% Typeset with `harvmac.tex', using `epsf' to include figure files,
% and the `xypic' package to typeset commutative diagrams.

\input harvmac.tex
%\draftmode
\noblackbox
\let\includefigures=\iftrue
\newfam\black
\includefigures

\input epsf
\def\plb#1 #2 {Phys. Lett. {\bf B#1} #2 }
\long\def\del#1\enddel{}
\long\def\new#1\endnew{{\bf #1}}
\let\<\langle \let\>\rangle 
\def\hbr{\hfil\break}
\def\figin{\epsfcheck\figin}\def\figins{\epsfcheck\figins}
\def\epsfcheck{\ifx\epsfbox\UnDeFiNeD
\message{(NO epsf.tex, FIGURES WILL BE IGNORED)}
\gdef\figin##1{\vskip2in}\gdef\figins##1{\hskip.5in}% blank space instead
\else\message{(FIGURES WILL BE INCLUDED)}%
\gdef\figin##1{##1}\gdef\figins##1{##1}\fi}
\def\DefWarn#1{}
\def\figinsert{\goodbreak\midinsert}
\def\ifig#1#2#3{\DefWarn#1\xdef#1{fig.~\the\figno}
\writedef{#1\leftbracket fig.\noexpand~\the\figno}%
\figinsert\figin{\centerline{#3}}\medskip
\centerline{\vbox{\baselineskip12pt
\advance\hsize by -1truein\noindent
\footnotefont{\bf Fig.~\the\figno:} #2}}
\bigskip\endinsert\global\advance\figno by1}
%%%
\else
\def\ifig#1#2#3{\xdef#1{fig.~\the\figno}
\writedef{#1\leftbracket fig.\noexpand~\the\figno}%
%\figinsert\figin{\centerline{#3}}\medskip
\centerline{\vbox{\baselineskip12pt
%\advance\hsize by -1truein\noindent
\footnotefont{\bf Fig.~\the\figno:} #2}}
%\bigskip\endinsert
\global\advance\figno by1}
\fi

\input xy
\xyoption{all}
\font\cmss=cmss10 \font\cmsss=cmss10 at 7pt
\def\inbar{\,\vrule height1.5ex width.4pt depth0pt}
\def\IC{{\relax\hbox{$\inbar\kern-.3em{\rm C}$}}}
\def\IP{{\relax{\rm I\kern-.18em P}}}
\def\IF{{\relax{\rm I\kern-.18em F}}}
\def\IZ{\relax\ifmmode\mathchoice
{\hbox{\cmss Z\kern-.4em Z}}{\hbox{\cmss Z\kern-.4em Z}}
{\lower.9pt\hbox{\cmsss Z\kern-.4em Z}}
{\lower1.2pt\hbox{\cmsss Z\kern-.4em Z}}\else{\cmss Z\kern-.4em
Z}\fi}
\def\IR{{\relax{\rm I\kern-.18em R}}}
\def\IQ{\relax\ifmmode\mathchoice
{\hbox{\cmss Q\kern-.4em Q}}{\hbox{\cmss Q\kern-.4em Q}}
{\lower.9pt\hbox{\cmsss Q\kern-.4em Q}}
{\lower1.2pt\hbox{\cmsss Q\kern-.4em Q}}\else{\cmss Q\kern-.4em
Q}\fi}

\def\cc{{\cal C}}
\def\co{{\cal O}}
\def\cl{{\cal L}}

\def\cm{{\cal M}}

\def\cv{{\cal V}}

\def\o0{\omega_0}

\def\w5{\omega_5}
\def\w6{\omega_6}
\def\O0{\Omega_0}

\def\ep{{\epsilon}}
\def\l{\lambda}
\Title{\vbox{\baselineskip12pt\hbox{hep-th/0104254}}}
{\vbox{
\centerline{D-Branes on Noncompact Calabi-Yau Manifolds:}
\smallskip
\centerline{K-Theory and Monodromy}}}
%\smallskip
{\vbox{\centerline {Xenia de la Ossa, Bogdan Florea and Harald Skarke}}}
\bigskip
\bigskip 
\medskip
\centerline{{ \it Mathematical Institute, University of Oxford,}}
\centerline{{\it 24-29 St. Giles', Oxford OX1 3LB, England}} 
\bigskip
\bigskip
\bigskip
\noindent
We study D-branes on smooth noncompact toric Calabi-Yau manifolds that
are resolutions of abelian orbifold singularities.
Such a space has a distinguished basis $\{S_i\}$ for the compactly 
supported K-theory. 
Using local mirror symmetry we demonstrate that the $S_i$ have simple
transformation properties under monodromy; in particular, they are the
objects that generate monodromy around the principal component of the
discriminant locus.
One of our examples, the toric resolution of
$\IC^3/(\IZ_2\times\IZ_2)$,  is a three parameter model for which we
are able to give an explicit solution of the GKZ system.

\Date{April 2001}

\def\atamp{{\it Adv. Theor. Math. Phys.}}

%%%%%%%%%%%%%%%%%%%%%%%%%%%%%%%%%%%%%
\nref\konts{M. Kontsevich, ``Homological Algebra of Mirror Symmetry'',
alg-geom/9411018.}
\nref\DK{E. Witten,``D-Branes and K-Theory'', {\it JHEP} 
{\bf 12} (1998) 019,  hep-th/9810188.}
%%%%%%%%%%%%%%%%%%%%%%%%%%%%%%%%%%%%%%%%%%%%%%%%%%%%%%%%%%%%%%%%%%%%%
\nref\paul{R. P. Horja, ``Hypergeometric functions and mirror symmetry
in toric varieties'', math.AG/9912109.}
\nref\psrpt{P. Seidel and R.P. Thomas, ``Braid group actions on
derived categories of coherent sheaves'', math.AG/0001043.}
\nref\thomass{R. Thomas, ``Mirror symmetry and actions of braid groups
on derived categories'', math.AG/0001044.}
\nref\hosono{S. Hosono, ``Local mirror symmetry and Type IIA monodromy
of Calabi-Yau manifolds'', hep-th/0007071.}
\nref\szend{B. Szendr\H oi, ``Diffeomorphisms and families of Fourier--Mukai
transforms in mirror symmetry'', math.AG/0103137.}
\nref\RPH{R. P. Horja, ``Derived category automorphisms from mirror
symmetry'', math.AG/0103231.}
%%%%%%%%%%%%%%%%%%%%%%%%%%%%%%%%%%%%%%%%%%%%%%%%%%%%%%%%%%%%%%%%%%%%%
\nref\dg{D.-E. Diaconescu and J. Gomis, ``Fractional branes and boundary
  states in orbifold theories'', {\it JHEP} {\bf 0010} (2000) 001, 
hep-th/9906242.}
\nref\GL{B. Greene and C. Lazaroiu, ``Collapsing D-Branes in 
Calabi-Yau Moduli Space: I'', hep-th/0001025.}%
\nref\DFR{M. R. Douglas, B. Fiol and C. R\"omelsberger,
``The spectrum of BPS branes on a noncompact Calabi-Yau'', 
hep-th/0003263.}%
\nref\FM{B. Fiol and M. Marino, ``BPS states and algebras from quivers'', 
{\it JHEP} {\bf 0007} (2000) 031, hep-th/0006189.}
\nref\ddd{D.-E. Diaconescu and M. R. Douglas, ``D-branes on stringy
Calabi-Yau manifolds'', hep-th/0006224.}
\nref\MOY{K. Mohri, Y. Onjo and S.-K. Yang, ``Closed Sub-Monodromy 
Problems, 
Local Mirror Symmetry and Branes on Orbifolds'', hep-th/0009072.}%
\nref\GJ{S. Govindarajan and T. Jayaraman, ``D-branes,
exceptional sheaves and quivers on Calabi-Yau manifolds: from Mukai to
McKay'', hep-th/0010196.}
\nref\AT{A. Tomasiello, ``D-branes on Calabi-Yau manifolds and
helices'', {\it JHEP} {\bf 0102} (2001) 008, hep-th/0010217.}
\nref\mayr{P. Mayr, ``Phases of supersymmetric D-branes on Kahler
manifolds and the McKay correspondence'', {\it JHEP} {\bf 0101} (2001)
018, hep-th/0010223.}
\nref\ACRY{B. Andreas, G. Curio, D. H. Ruiperez and S.-T. Yau, 
``Fourier-Mukai transform and mirror symmetry for D-branes on elliptic
Calabi-Yau'', math.AG/0012196.}
\nref\LMW{W. Lerche, P. Mayr and J. Walcher, ``A new kind of McKay
correspondence from non-abelian gauge theories'', hep-th/0103114.}
\nref\aspmon{P. S. Aspinwall, ``Some Navigation Rules for D-Brane
Monodromy'', hep-th/0102198.}
%%%%%%%%%%%%%%%%%%%%%%%%%%%%%%%%%%%%%%%%%%%%%%%%%%%%%%%%%%%%%%%%%%%%%%%%
\nref\mckay{J. McKay, ``Graphs, singularities and finite groups'',
Proc. Symp. in Pure Math. {\bf 37} (1980), 183.}
\nref\gsv{G. Gonzales-Sprinberg and J.-L. Verdier, ``Construction
geometrique de la correspondance de McKay'', Ann. sci. ENS {\bf 16}
(1983), 409.}
\nref\reidmk{M. Reid, ``McKay correspondence'', alg-geom/9702016.}
\nref\nakamura{I. Nakamura, ``Hilbert schemes of Abelian group
orbits'', preprint.}
\nref\itonakajima{Y. Ito and H. Nakajima, ``McKay correspondence and
  Hilbert schemes in dimension three'', math.AG/9803120.}
\nref\BKR{T. Bridgeland, A. King, and M. Reid, ``Mukai implies
McKay'', math.AG/9908027.}
\nref\crawreid{A. Craw and M. Reid, ``How to calculate A-Hilb
$\IC^3$'', math.AG/9909085.}
\nref\craw{A. Craw, ``An explicit construction of the McKay
correspondence for A-Hilb $\IC^3$'', math.AG/0010053.} 
%%%%%%%%%%%%%%%%%%%%%%%%%%%%%%%%%%%%%%%%%%%%%%%%%%%%%%%%%%%%%%%%%%%%%%%%%
\nref\ful{W. Fulton, ``Introduction to Toric Varieties'', 
        Princeton Univ. Press, Princeton 1993.}
\nref\oda{T. Oda, ``Convex Bodies and Algebraic Geometry'', Springer, Berlin
        Heidelberg 1988.}
\nref\cox{D. Cox, ``The Homogeneous Coordinate Ring of a Toric Variety'',
        J. Alg. Geom. {\bf 4} (1995) 17, alg-geom/9210008.}
\nref\coka{D. Cox and S. Katz, ``Mirror Symmetry and Algebraic
        Geometry'', American Mathematical Society, 1999.}
%%%%%%%%%%%%%%%%%%%%%%%%%%%%%%%%%%%%%%%%%%%%%%%%%%%%%%%%%%%%%%%%%%%%%%%%%%
\nref\witten{E.~Witten, ``Phases of $N=2$ theories in two
dimensions'', {\it Nucl. Phys.} {\bf B403} (1993) 159, hep-th/9301042.}
\nref\agm{P. S. Aspinwall, B. R. Greene, and D. R. Morrison, ``Multiple Mirror 
        Manifolds and Topology Change in String Theory'', \plb 303 (1993) 249,
        hep-th/9301043.}
\nref\ckyz{T.-M. Chiang, A. Klemm, S.-T. Yau, and E. Zaslow, ``Local
  Mirror Symmetry: Calculations and Interpretations'', \atamp\ {\bf 3}
  (1999), 495, hep-th/9903053.}
\nref\agmd{P. S. Aspinwall, B. R. Greene, and D. R. Morrison,
``Measuring small distances in $N=2$ sigma models'', {\it Nucl. Phys.} 
{\bf B420} (1994) 184, hep-th/9311042.}
%%%%%%%%%%%%%%%%%%%%%%%%%%%%%%%%%%%%%%%%%%%%%%%%%%%%%%%%%%%%%%%%%%%%%%%%%
\nref\categ{M. R. Douglas, ``D-branes, categories and ${\cal N}=1$
supersymmetry'', hep-th/0011017.}
\nref\AL{P. S. Aspinwall and A. Lawrence, ``Derived categories and
zero-brane stability'', hep-th/0104147.}
\nref\CIL{C. I. Lazaroiu, ``Generalized complexes and string field
theory'', hep-th/0102122.}
\nref\DED{D.-E. Diaconescu, ``Enhanced D-brane categories from string
field theory'', hep-th/0104200.}
\nref\fultoni{W. Fulton, ``Intersection Theory'', Springer-Verlag (1984).}
\nref\dgm{M. R. Douglas, B. R. Greene, and D. R. Morrison, ``Orbifold
  resolution by D-branes,'' {\it Nucl. Phys.} {\bf B506} (1997) 84,
  hep-th/9704151.}
\nref\bfmp{P. Baum, W. Fulton, and R. MacPherson, ``Riemann-Roch and
topological K-theory for singular varieties'', {\it Acta Mathematica}
{\bf 143} (1979) 155.}
\nref\iversen{B. Iversen, ``Local Chern classes'', {\it
Ann. scient. Ec. Norm. Sup.} {\bf t. 9} (1976) 155.} 
\nref\dr{D.-E. Diaconescu and C. R\"omelsberger, ``D-branes and 
  bundles on elliptic fibrations'', {\it Nucl. Phys.} {\bf B574} (2000) 
  245, hep-th/9910172.}%
\nref\aspi{P. S. Aspinwall, ``Resolution of orbifold singularities in
string theory'', hep-th/9403123.} 
\nref\zaslow{E. Zaslow, ``Solitons and helices: The search for a
math-physics bridge'', {\it Commun. Math. Phys.} {\bf 175} (1996) 337, 
hep-th/9408133.}
\nref\VHI{K. Hori, A. Iqbal and C. Vafa, ``D-Branes And Mirror Symmetry'', 
hep-th/0005247.}%
\nref\kvino{A. B. Kvichansky and D. Yu. Nogin, ``Exceptional
collections on ruled surfaces'', in {\it Helices and Vector Bundles:
 Seminaire Rudakov}, A. N. Rudakov et al, {\it LMS Lecture Note
 Series} {\bf 148} 97.} 
\nref\BRG{B. R. Greene, ``D-brane topology changing transitions'', 
{\it Nucl. Phys.} {\bf B525} (1998) 284, hep-th/9711124.}%
\nref\AP{P. S. Aspinwall and M. R. Plesser, ``D-branes, discrete
torsion and the McKay correspondence'', {\it JHEP} {\bf 0102} (2001)
009, hep-th/0009042.} 
\nref\mukray{S. Mukhopadhyay and K. Ray, ``Fractional Branes on a
Non-compact Orbifold'', hep-th/0102146.}
%
%%%%%%%%%%%%%%%%%%%%%%%%%%%%%
%

\newsec{Introduction}

The simplest manifestation of mirror symmetry is an exchange of the
Hodge numbers $h_{ii}$ and $h_{i,n-i}$ of a Calabi-Yau $n$-fold $X$
and its mirror $\tilde X$.
Interpreted naively, this would seem to imply identifications between
$n$-cycles on $\tilde X$ and holomorphic cycles on $X$. 
This leads to the following puzzle.
Monodromy in the complex structure moduli space of $\tilde X$ can take 
$n$-cycles to arbitrary other $n$-cycles, so this would lead to the
counterintuitive picture of mixing cycles of arbitrary even 
dimension on $X$.

Mathematically this puzzle is resolved by Kontsevich's conjecture
\konts\ that the relevant objects on $X$ are the elements of the bounded 
derived category $D^b$ of coherent sheaves.
In terms of physics, we now have the following intuitive picture.
We should not think of a cycle as a geometric object per se, but
as something that a D-brane can wrap.
A D-brane corresponds to a cycle with a vector bundle on it only in
a semiclassical limit.
In a more general construction a D-brane can be obtained from
higher dimensional branes and anti-branes, leading to an
interpretation in terms of K-theory \DK\ that is consistent with
Kontsevich's approach.

Monodromy in the complexified K\"ahler moduli space of Calabi-Yau
manifolds has been the object of recent studies both by mathematicians
\paul -\RPH\ and by physicists \dg -\aspmon .
One particular approach \ddd\GJ\AT\mayr\ uses well known results on McKay
correspondence \mckay -\craw\ to obtain a special basis
for the K-theory on $X$. 
These authors study noncompact toric Calabi-Yau manifolds that are
resolutions of singularities of the type $\IC^d/\IZ_n$ (or more
general Calabi-Yau singularities in \LMW ) with a single
exceptional divisor, mainly in order to describe compact Calabi-Yau
manifolds as hypersurfaces in the exceptional divisor.

In this work we study D-branes on non-compact toric Calabi-Yau
manifolds in their own right, with the aim of getting a better
understanding of what the fundamental D-brane degrees of freedom are
and how they behave under monodromy.
We show how to construct a distinguished basis for the compactly
supported K-theory with a number of remarkable properties, the most
striking being the fact that the elements of this basis seem to
generate the monodromy around the principal component of the
discriminant locus in the same way as the structure sheaf $\co_X$ does 
in the compact case. We consider cases with 
more than one exceptional divisor, and we test the applicability
of the above statements beyond the realm of McKay correspondence.
We do not have general proofs for our statements, but we
demonstrate their validity in various examples with the help of local
mirror symmetry.

The outline of this paper is as follows.
In the next section we present necessary material on toric varieties,
their Mori and K\"ahler cones, and the secondary fan.
In section 3 we introduce local mirror symmetry, toric moduli spaces
and the GKZ system.
While the material in these sections is known, its presentation
relying on the holomorphic quotient approach to toric varieties may be
useful; besides, it serves to establish notations and to introduce
some of our examples.
Section 4 is the core of this paper.
There we discuss K-theory and known results related to McKay
correspondence and proceed to define %our own construction of
the distinguished generators $S_i$ of the compactly supported K-theory.
We find that these generators are the ones that are responsible for
monodromy around the principal component of the discriminant locus.
In section 5 we demonstrate that our methods work in cases that are
more complicated than examples of the type $\IC^d/\IZ_n$.
We consider the case of
$\IC^3/(\IZ_2\times\IZ_2)$ and find that it is possible to solve the
corresponding GKZ system, with results that agree precisely with our
assertions.

\newsec{Toric Calabi-Yau manifolds}

We start with presenting some general considerations on non-compact
toric Calabi-Yau manifolds and their K\"ahler and Mori cones that will
be useful later. 
The results obtained here are standard \ful -\coka , but
our derivations from basic facts in toric geometry are possibly
simpler than what can be found in the literature.

The data of a $d$-dimensional toric variety $X$ can be specified
in terms of a fan $\Sigma$ in a lattice $N$ isomorphic to $\IZ^d$.
$X$ is smooth whenever each of the $d$-dimensional cones in
$\Sigma$ is generated over $\IR_+$ by exactly $d$ lattice vectors that  
generate $N$ over $\IZ$.
We will only consider this case.

Perhaps the simplest way of describing $X$ is as follows:
Assume that there are $k$ one dimensional cones in $\Sigma$ generated
by lattice vectors $v_1,\ldots,v_k$.
Assign a homogeneous variable $z_i$ to each of the $v_i$ and a
multiplicative equivalence relation among the $z_i$, 
\eqn\multrel{(z_1,\ldots,z_k)\sim (\l^{q_1}z_1,\ldots,\l^{q_k}z_k)} 
with $\l\in\IC^*$ for any linear relation $q_1 v_1+\cdots +q_kv_k=0$
among the generators $v_i$.
The $q_i$ can be normalized to be integers without common divisor;
in the context of a gauged linear sigma model they are the charges
with respect to the $U(1)$ fields.
The number of independent relations of the type \multrel\ is $k-d$.

Define a subset $\IC^k\setminus F_\Sigma$ of
$\IC^k=\{(z_1,\ldots,z_k)\}$ as the set of all $k$-tuples of $z_i$
with the following property:
If $z_i$ vanishes for all $i\in I\subset\{1,\ldots,k\}$, then all
$v_i$ with $i\in I$ belong to the same cone.
Then $X$ is $(\IC^k\setminus F_\Sigma)/(\IC^*)^{k-d}$, where the
division by $(\IC^*)^{k-d}$ is implemented by taking equivalence
classes with respect to the multiplicative relations \multrel.

Every one dimensional cone generated by $v_i$ corresponds in a natural
way to the divisor $D_i$ determined by $z_i=0$.
Similarly, an $l$ dimensional cone spanned by $v_{i_1},\ldots,v_{i_l}$
determines the codimension $l$ subspace $z_{i_1} =\ldots = z_{i_l}=0$
of $X$.

Monomials of the type $z_1^{a_1}\ldots z_k^{a_k}$ are sections of line
bundles $\co(a_1 D_1+\cdots + a_k D_k)$. 
If we denote by $M$ the lattice dual to $N$ and by $\<~~,~~\>$ the
pairing between $N$ and $M$, it is easily checked that monomials of
the form $z_1^{\<v_1,m\>}\ldots z_k^{\<v_k,m\>}$ with $m\in M$ are
meromorphic functions (i.e., invariant under \multrel) on $X$.
This implies the linear equivalence relations 
\eqn\lineq{
\<v_1,m\>D_1+\ldots+\<v_k,m\>D_k\sim 0~~~\hbox{for any}~~m\in M.}
Conversely, if a divisor of the form $a_1 D_1+\cdots a_k D_k$ belongs
to the trivial class, then there exists an $m\in M$ such that
$a_i=\<v_i,m\>$ for all $i$.

A calculation similar to the way the canonical divisor of $\IP^d$ is
determined shows that the canonical divisor of $X$ is given by 
$-D_1-\cdots-D_k$.
Thus $X$ is Calabi-Yau if and only if $D_1+\cdots+D_k$ is trivial,
i.e. if and only if there exists an $m\in M$ such that $\<v_i,m\>=1$
for every $i$.
Therefore the $v_i$ must all lie in the same affine hyperplane. 
We will make use of this fact by drawing toric diagrams in dimension
$d-1$ that display only the endpoints of the $v_i$.

We will be interested in the K\"ahler moduli space of $X$.
The dual of the K\"ahler cone is the Mori cone spanned by effective
curves. 
Toric curves are determined by $(d-1)$-dimensional cones
$\sigma_{d-1}$ in $\Sigma$.
If a curve is compact, the corresponding cone is the boundary between
two $d$-dimensional cones $\sigma_d^{(1)}$, $\sigma_d^{(2)}$.
If we denote the integer generators of $\sigma_{d-1}$,
$\sigma_d^{(1)}$, $\sigma_d^{(2)}$ by $\{v_1,\ldots,v_{d-1}\}$,
$\{v_1,\ldots,v_{d-1}, v_d\}$, $\{v_1,\ldots,v_{d-1}, v_{d+1}\}$,
respectively (remember that we are assuming that our cones are
simplicial and their generators generate $N$), we find that
$v_d+v_{d+1}$ must lie in the intersection of the hyperplane of 
$\sigma_{d-1}$ with $N$ and so there exists a unique
linear relation of the form $l_1v_1+\ldots +l_{d+1}v_{d+1}=0$ with
$l_d=l_{d+1}=1$ and all $l_i$ integer.

We will now argue that the $l_i$ are actually the intersection numbers
between the curve $C=D_1\cdot\ldots\cdot D_{d-1}$ determined by 
$\sigma_{d-1}$ and the toric divisors $D_i$.
Our general rules imply that intersection numbers between $d$ different
toric divisors are 1 or 0 depending on whether these divisors form a
cone in $\Sigma$.
This implies $C\cdot D_d=l_d=1$, $C\cdot D_{d+1}=l_{d+1}=1$ and
$C\cdot D_i=0$ for $i>d+1$.
For calculating $C\cdot D_i$ with $i<d$ we have to use linear
equivalence relations of the type \lineq .
To calculate $C\cdot D_1$ we may choose $m$ to fulfill $\<v_1,m\>=1$
and $\<v_2,m\>=\cdots=\<v_d,m\>=0$.
Then 
\eqn\CD{\eqalign{
0\sim \sum \<v_i,m\>D_i=&\<v_1,m\>D_1+\<v_{d+1},m\>D_{d+1}+\ldots =\cr
&D_1+\<-l_1v_1-\cdots -l_dv_d,m\>D_{d+1}+\ldots=
D_1-l_1D_{d+1}+\ldots,}}
i.e. $D_1\sim l_1D_{d+1}+\ldots$ where `$\ldots$' stands for $D_i$
with $i>d+1$ which do not intersect $C$.
Thus we find that $C\cdot D_1=l_1C\cdot D_{d+1}=l_1$.
As our choice of $D_1$ among the $D_i$ with $i<d$ was arbitrary, we
have indeed shown that $C\cdot D_i=l_i$ for any $i$.

A set of generators for the Mori cone is then given by all those
curves $C^{(i)}$ whose $l^{(i)}$ cannot be written as nonnegative linear
combinations of the other $l^{(j)}$.
The matrix $L$ whose lines are the $l^{(i)}$ of the Mori cone
generators has the following remarkable properties:
Any Matrix $Q$ consisting of $d-k$ independent (linear combinations
of) lines of $L$ serves as a `charge matrix' for the relations 
\multrel\ .
If the Mori cone is simplicial, we just have $L=Q$.
This will be the case in most of our examples, so we will not
distinguish between $L$ and $Q$ in these cases.
Any column of $L$ is associated with a toric divisor $D_i$.
If a linear combination $\sum_jL_{ij}a_j$ of column vectors of $Q$
vanishes, then the corresponding divisor $\sum_ja_jD_j$ has vanishing
intersection with any effective curve, i.e. it is trivial.
Therefore a diagram displaying the column vectors of $L$ or $Q$
encodes the linear equivalence relations among the toric divisors $D_i$. 
We may interpret these vectors as one dimensional cones of a fan, the
so called `secondary fan' of $X$.
Note, however, that two distinct but linearly equivalent toric
divisors correspond to the same vector in the secondary fan.
As the entries of $L$ are the intersections between the
generators of the Mori cone and the divisors, the K\"ahler cone of $X$
is determined by those $\sum_ja_jD_j$ such that the corresponding
linear combinations of the columns of $L$ only have nonnegative
entries.

We should stress that our analysis was in terms of a single fixed
triangulation. 
If we allow several distinct triangulations, the Mori cone vectors of
any of them will lead to correct charge matrices $Q$ but the K\"ahler
condition will depend on which combinations of the charge vectors
correspond to the Mori cone, i.e. on the choice of triangulation.
In this way several regions of a secondary fan constructed from some
charge matrix $Q$ can correspond to different `geometric phases' in
the sense of \witten\agm.

We will now present some of the examples that we are going to use in this
paper. 
\del
These examples are chosen with the aim of relating our
constructions to known or conjectured facts; 
in section 5 we will present further examples that go beyond cases
where McKay correspondence applies.
\enddel

\noindent {\bf Example 1:}\hbr
{\it The toric resolution of}~~ $\IC^2/\IZ_n$:
We have toric divisors $D_0,\ldots,D_n$ corresponding to vectors 
\eqn\vctwo{
v_0=\left({0\atop 1}\right),~~v_1=\left({1\atop 1}\right),~~\cdots,
~~v_n=\left({n\atop 1}\right).}
$D_0$ and $D_n$ are non-compact and correspond to the coordinates
of the original $\IC^2$ on which $\IZ_n$ acts by 
$(z_0,z_n)\to (\ep z_0,\ep^{n-1}z_n)$ with $\ep=e^{2\pi i/n}$.
All other $D_i$ are compact and are nothing but the effective curves.
The Mori cone vectors are determined by $v_{i-1}-2v_i+v_{i+1}=0$,
leading to 
\eqn\lctwo{Q=\left(\matrix{
1 & -2 & 1 & 0 & \ldots & 0 & 0 \cr
0 &  1 & -2 & 1 &\ldots & 0 & 0 \cr
 & & & &     \ldots & &         \cr
0 & 0 & 0 & 0 & \ldots & -2 & 1 }\right).}
Upon dropping the first and the last column, this becomes
$-M_{SU(n)}$, where $M_{SU(n)}$ is the Cartan matrix of $SU(n)$.
Thus the generators of the K\"ahler cone, corresponding to linear
combinations of the $D_i$ that turn the columns of $L$ into unit
vectors, are given by
$-\sum_{j=1}^{n-1}(M_{SU(n)})^{-1}_{ij}D_j$ or, alternatively, by
\eqn\kahlctwo{D_0,~~ D_1+2D_0~~, D_2+2D_1+3D_0,~~ \ldots,~~ 
D_{n-2}+2D_{n-3}+\cdots +(n-1)D_0. }

\noindent {\bf Example 2:}\hbr
{\it The toric resolution of}~ $\IC^n/\IZ_n$: 
The resolution of a singular space of the type $\IC^n/\IZ_n$, where
$\IZ_n$ acts on the coordinates of $\IC^n$ by
\eqn\act{
(z_1,\ldots,z_n)\to(\ep z_1,\ldots, \ep z_n) \hbox{ with } 
\ep=e^{2\pi i/n}}
can be represented torically by vectors $v_1,\ldots,v_{n+1}$ subject
to the single relation $v_1+v_2+\cdots +v_n=nv_{n+1}$;
the $N$ lattice is just the lattice generated by the $v_i$.
The first $n$ vectors $v_1,\ldots,v_n$ correspond to the original
coordinates $z_i$ whereas $v_{n+1}$ corresponds to the single
exceptional divisor $D_{n+1}=\{z_{n+1}=0\}$ isomorphic to $\IP^{n-1}$.
The Mori cone is determined by the single relation, leading to
\eqn\lcn{Q=(1,1,\ldots,1,-n).}

\noindent {\bf Example 3:}\hbr
{\it The toric resolution of}~ $\IC^3/\IZ_5$: 
We first consider a singular space of the type $\IC^3/\IZ_5$, where
$\IZ_5$ acts on the coordinates of $\IC^3$ by
\eqn\act{
(z_1,z_2,z_3)\to(\ep z_1,\ep^3 z_2, \ep z_3) \hbox{ with } 
\ep=e^{2\pi i/5}.}
As a toric variety $\IC^3/\IZ_5$ is determined by three vectors
\eqn\Frays{
v_1=\left(\matrix{-1\cr 0\cr 1\cr}\right),\qquad
v_2=\left(\matrix{2\cr 2\cr 1\cr}\right),\qquad
v_3=\left(\matrix{0\cr -1\cr 1\cr}\right)}
in a lattice $N$ isomorphic to $\IZ^3$, the singularity resulting
from the fact that $v_1$, $v_2$ and $v_3$ generate only a
sublattice of $N$.
A complete crepant (i.e., canonical class preserving) toric resolution
$X\to\IC^3/\IZ_5$ is obtained by adding two further rays
\eqn\Drays{
v_4=\left(\matrix{0\cr 0\cr 1\cr}\right),\qquad
v_5=\left(\matrix{1\cr 1\cr 1\cr}\right)}
and triangulating the resulting diagram (this triangulation is
unique in the present case).

\ifig\fan{The resolution of $\IC^3/\IZ_5$.}
{\epsfxsize2in\epsfbox{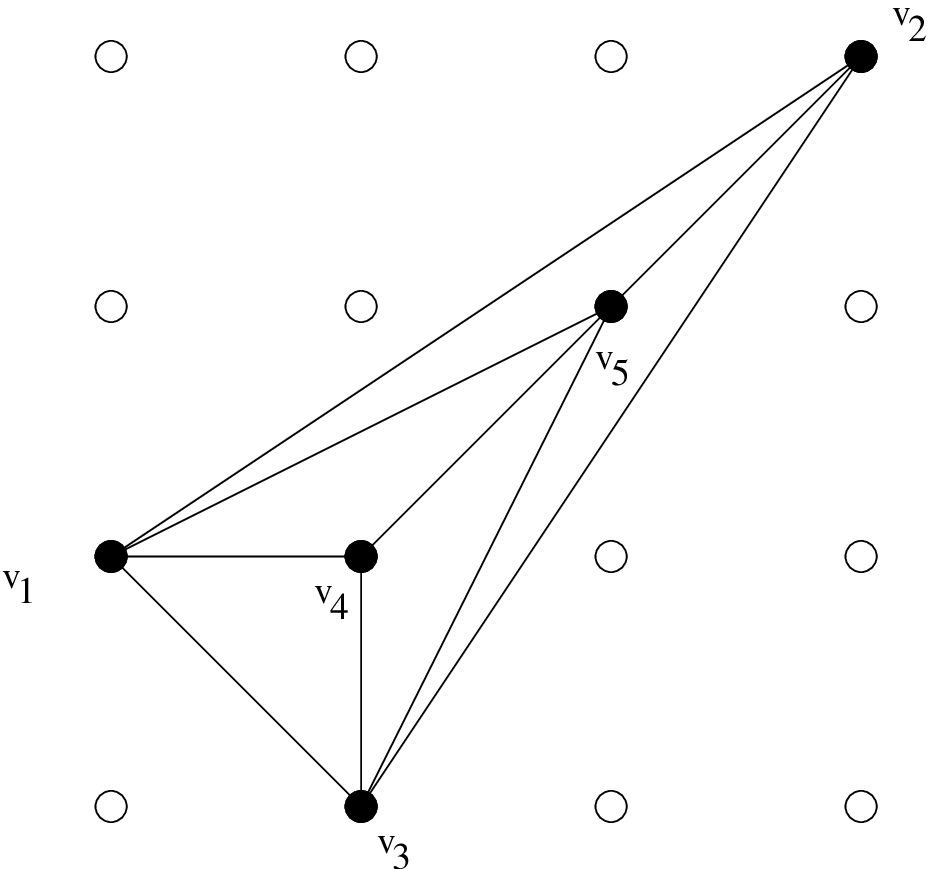}}

The resulting fan, with the redundant third coordinate suppressed, is 
shown in \fan.
The structure of the resolution is easily read off from this diagram:
We have two exceptional divisors $D_4$ and $D_5$ corresponding to
$v_4$ and $v_5$, respectively.
The star fans of $v_4$ and $v_5$ tell us that $D_4$ is a $\IP^2$
and $D_5$ is a Hirzebruch surface $\IF_3$.
$D_4$ and $D_5$ intersect along a curve $h$ which is a hyperplane
of the $\IP^2$ and at the same time the negative section of $\IF_3$. 
We denote by $D_1$, $D_2$ and $D_3$ the noncompact toric divisors 
corresponding to the vertices $v_1$, $v_2$ and $v_3$ respectively 
(i.e., the zero loci of the coordinates of our original $\IC^3$).
Intersection numbers can be calculated by using the linear
equivalences 
\eqn\rateq{
D_1\sim D_3\sim D_5+2D_2~~\hbox{ and }~~D_1+D_2+D_3+D_4+D_5\sim 0}
and the fact that three distinct toric divisors have an intersection
number of 1 if they belong to the same cone and 0 otherwise.
As $D_1$ is linearly equivalent to $D_3$ we omit expressions
involving $D_3$ in the following.
Intersections of divisors are well defined whenever they involve at
least one of $D_4$ and $D_5$.
Triple intersections are given by
\eqn\inters{\eqalign{
& D_4^3=9,~~D_4^2\cdot D_5=-3,~~D_4\cdot D_5^2=1,~~D_5^3=8, \cr
& D_4^2\cdot D_1=-3,~~D_4\cdot D_5\cdot D_1=1,~~D_5^2\cdot D_1=-2,
~~D_5^2\cdot D_2=-5, \cr
& D_4\cdot D_1^2=1,~~D_5\cdot D_1^2=0,~~D_5\cdot D_1\cdot D_2=1,
~~D_5\cdot D_2^2=3 \cr
}}
and the vanishing of $D_4\cdot D_2=0$.
Intersections of two distinct divisors are determined by
\eqn\twoint{
D_4\cdot D_5=D_4\cdot D_1=h,~~D_4\cdot D_2=0,~~D_5\cdot D_1=f,~~ 
D_5\cdot D_2=h+3f}
where $f$ is the fibre of the $\IF_3$.
The self-intersections of $D_4$ and $D_5$ are 
\eqn\selfint{D_4^2=-3h,~~~D_5^2=-2h-5f.}
We also have:
\eqn\intersb{\eqalign{
& D_4\cdot h=-3,~~D_4\cdot f=1,~~D_5\cdot h=1,~~D_5\cdot f=-2, \cr
& D_1\cdot h=1,~~D_1\cdot f=0,~~D_2\cdot h=0,~~D_2\cdot f=1.
}}
This implies that $(C_1,C_2)=(h,f)$ and $(D_1,D_2)$ form mutually dual bases of
the Mori cone and the K\"ahler cone of $X$.
In terms of codimension one (here, two dimensional) cones $\sigma$ 
and the linear relations between the rays in the two cones of maximal 
dimension that contain $\sigma$, we obtain the following linear
relations among the vectors $v_1,\ldots,v_5$ of the fan:
\eqn\chvect{\eqalign{
l^{(1)}=(1, 0, 1, -3, 1),\cr
l^{(2)}=(0, 1, 0, 1, -2).\cr}}
As a check on our intersection numbers, we observe that indeed 
$D_i\cdot C_j=l^{(j)}_i$.

\ifig\secfan{The secondary fan of the resolution of $\IC^3/\IZ_5$.}
{\epsfxsize2.5in\epsfbox{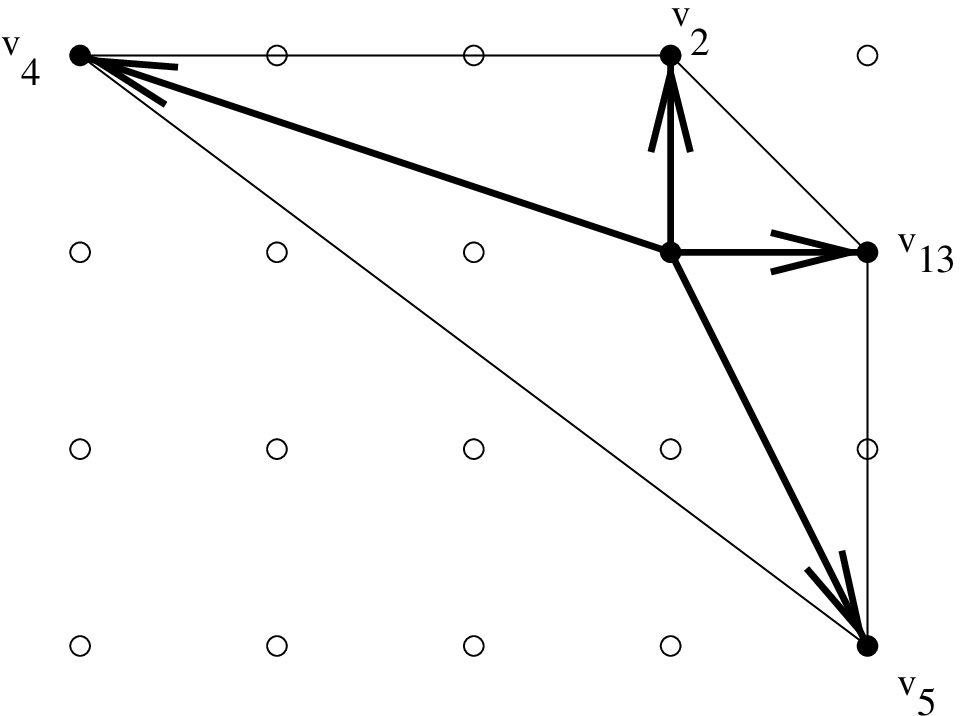}}

If we consider the
matrix whose lines are the generators \chvect\ of the Mori cone and
draw the rays corresponding to the columns of this matrix, we obtain
the secondary fan for $X$ as shown in \secfan.
The linear relations among the vectors in this fan encode the linear
equivalences \rateq\ among the divisors. 

\newsec{Local mirror symmetry}

In our study of D-brane states we will have to address issues that
involve quantum geometry.
A standard tool for this problem is the use of mirror symmetry.
In particular, classical periods in the mirror geometry get mapped to
quantum corrected expressions related to the middle cohomology of the
original space. 
In the non-compact case one has to use local mirror symmetry.
For our applications of this subject we have relied mainly on \ckyz\
and we refer to this paper for further references.
The authors of \ckyz\ consider decompactifications of Calabi-Yau 
hypersurfaces in toric varieties such that the volumes of certain 
cycles remain compact.
They show that in the decompactification limit these cycles lead to 
differential equations that are identical with the GKZ differential
systems of a lower dimensional geometry. 
We will assume that this remains true even for cases where the
non-compact Calabi-Yau geometry cannot be identified with a limiting
case of a compact Calabi-Yau hypersurface.

The local mirror of a $d$-dimensional noncompact Calabi-Yau geometry is 
determined by interpreting the diagram of the hyperplane containing the
end points of the $v_i$ now as a polytope $P$ in a $(d-1)$-dimensional 
lattice $\tilde M$.
A polytope corresponds to a line bundle $\cl$ over a toric variety
$\cv$ by the following construction:
Fix any point in $\tilde M$ to be the origin.
Describe the facets of $P$ by equations 
$E_j(\tilde m):=\< \tilde v_j, \tilde m \> +c_j = 0$,
where $\tilde v_j\in\tilde N$, the lattice dual to $\tilde M$ and fix
the sign ambiguity about $\tilde v_j$ in such a way that 
$E_j(\tilde m)$ is nonnegative for points $\tilde m$ of $P$.
Choose $\cv$ to be a toric variety whose one dimensional rays are the
$\tilde v_j \in \tilde N$ corresponding to a variable $x_j$ as
in the previous section.
To every point $\tilde m \in \tilde M$ assign the monomial 
$\prod_j x_j^{E_j(\tilde m)}$.
Then $\cl$ is the bundle whose sections are determined by polynomials
of the type
\eqn\locmir{P(a;x)=\sum_{i=1}^k a_i \prod_j x_j^{E_j(\tilde m_i)}.}
The `local mirror' $\tilde X$ of $X$ is defined to be the vanishing
locus of a section \locmir\ of $\cl$.

In the present context we can give an alternative description of the
$E_j$: 
We have $N\simeq \tilde M\oplus \IZ$ and may choose coordinates such that
$v_i=(\tilde m_i,1)$.
Then we can write the affine function $E_j(\tilde m_i)$ as a linear
function of the form $\<v_i, \tilde v_j' \>$ with 
$\tilde v_j'\in\hbox{Hom}(N,\IZ)=M$
(it is easy to check that the $\tilde v_j'$ are the elements of $M$ dual to
the $(d-1)$-dimensional cones at the boundary of the support of
$\Sigma$).

Obviously the complex structure moduli space of $\tilde X$ is
parametrized by the $a_i$. 
It is important to note, however, that different sets of $a_i$ need
not correspond to different complex structures.
In particular, a scaling $x_j\to \l_j x_j$ does not amount to a change
in the complex structure but leads to a redefinition of the $a_i$, 
implying the equivalences
\eqn\modspa{\eqalign{(a_1,a_2,\ldots, a_k)\sim 
&(\l_j^{E_j(\tilde m_1)}a_1, \l_j^{E_j(\tilde m_2)}a_2,
\ldots, \l_j^{E_j(\tilde m_k)}a_k)=\cr
&(\l_j^{\<v_1, \tilde v_j'\>}a_1, \l_j^{\<v_2, \tilde v_j'\>}a_2,
\ldots,  \l_j^{\<v_k, \tilde v_j'\>}a_k)}}
for any $j$.
Given identifications of this type it is natural to seek a
description in terms of toric geometry. 
If we interpret the exponents of the $\l$'s as linear relations among
vectors $u_i$ in a toric diagram and notice that the $\tilde v_j'$ generate 
$M$ (at least over the rational numbers), we find that the $u_i$ fulfill 
\eqn\linru{\<m, v_1\>u_1+\<m, v_2\>u_2+\cdots +\<m,v_k\>u_k=0}
for any $m\in M$.
These are just the relations among the vectors of the secondary fan
which encodes, as we saw, the linear equivalence relations \lineq\ of
the divisors $D_i$ corresponding to the $v_i$.
There are some subtleties, however: 
As we saw in the previous section, it is possible that two distinct 
(but linearly equivalent) toric divisors lead to the same vector in
the secondary fan. 
We will show how to interpret this in the context of the examples.
Besides, it is possible that there are identifications in the moduli
space that do not come from rescalings of the type $x_j\to \l_j x_j$
and hence have a structure different from \modspa .
If this occurs, the toric variety associated with the secondary fan is
called the `simplified moduli space' $\cal M_{\rm simp}$.
Depending on whether we have extra identifications or not, the toric
variety corresponding to the secondary fan is a compactification of
$\cal M_{\rm smooth}$ (the moduli space of all smooth local mirror 
hypersurfaces) or a covering space of a compactification of 
$\cal M_{\rm smooth}$.

$\tilde X$ will degenerate over various loci in $\cal M_{\rm simp}$
where $\partial P(a;x)/\partial x_j=0$ can be solved for all $j$
without violating the conditions on which $x_i$ are allowed to vanish
simultaneously. 
Some of these loci may just be toric divisors, but usually there is
also at least one connected piece given by a polynomial equation in
the $a_i$ to which we will refer as the primary or principal
component of the discriminant locus.

If we want to relate the mirror geometry to the original one, we have
to find a region in the moduli space where quantum corrections are
strongly suppressed.
This is the case for the deep interior of the K\"ahler cone, the so
called large volume limit, which is dual to the large complex
structure limit.
As we saw in section 2, the K\"ahler cone can be determined by writing
any divisor as a linear combination of toric divisors and demanding
that the corresponding linear combination of columns of the matrix $L$
contain only nonnegative entries.
If the resulting generators do not belong to the secondary fan, we
have to blow up the moduli space in order to be able to change to the
large complex structure variables.

In those cases where the Mori cone is simplicial we can draw the
secondary fan by displaying the columns of $L$ and the generators of
the K\"ahler cone will be nothing but the unit vectors.
If we then write the linear relations among the vectors in the
secondary fan in such a way that we express every vector in
terms of the unit vectors and use the corresponding rules \multrel\
to set all variables except the large complex structure 
variables\foot{We hope that no confusion arises from the fact that we
use the same symbol $z_i$ for the coordinates of $X$ and the large complex
structure variables.}
$z_i$ to 1, we find that the $z_i$ can be expressed as 
\eqn\lcssimp{z_i=\prod_{j=1}^k a_j^{l^{(i)}_j}.}
Note that we do not include a sign here (compare with e.g. \agmd).

If $X$ is the resolution of an orbifold singularity of the type
$\IC^d/\IZ_n$ there is another distinguished coordinate patch in the
moduli space containing the orbifold locus where all $a_i$ except the
ones corresponding to the coordinates of the $\IC^d$ are set to zero.
At this point the conformal field theory is expected to acquire a
quantum symmetry.
We find that the moduli space in this case always has a singularity
that looks locally like $\IC^{{\rm dim}~\cm}/\IZ_n$.

The GKZ differential operators are calculated by using the following
recipe:
For every linear relation $\sum l_jv_j=0$,  where $l$ corresponds
to {\it any curve in the Mori cone} (see \ckyz) we define a
differential operator in terms of the $a_i$,
\eqn\gkzgen{{\cal D}=\prod_{j:~l_j>0}\partial_{a_j}^{l_j}-
   \prod_{j:~l_j<0}\partial_{a_j}^{-l_j}.}
Assume that we work in a specific coordinate patch given by some
$\phi_i=\prod a_j^{\mu_{ij}}$.
In order to transform \gkzgen\ to a system involving the $\phi_i$ we
can rewrite it in terms of operators 
$\Theta_{a_j}:=a_j\partial_{a_j}$, commute all $a_j$ to
the left using $\Theta_{a_j} a_j^{-1}= a_j^{-1}(\Theta_{a_j}-1)$ and then
express the $\Theta_{a_i}$ as $\sum_i \mu_{ij}\Theta_{\phi_i}$ with
$\Theta_{\phi_i}:=\phi_i\partial_{\phi_i}$.

We stress that the solutions of the GKZ system are not the periods on
$\tilde X$ but rather the logarithmic integrals of the periods.
While the periods are finite and non-vanishing on the moduli space
wherever $\tilde X$ is non-degenerate, the GKZ solutions have extra
singularities at the zero loci of moduli space coordinates coming from
the logarithmic integration. 
The GKZ solutions are multivalued and undergo monodromy
transformations around codimension one loci where they are not
holomorphic.
We will be interested mainly in monodromies around the large complex
structure divisors $z_i=0$ and around the principal component of the
discriminant locus.
In addition, there is the possibility of a non-trivial transformation
(`orbifold monodromy', which, strictly speaking, is not a monodromy)
if the moduli space looks locally like $\IC^{{\rm dim}~\cm}/\IZ_n$.

We will now show how these concepts can be applied to our examples.

\noindent {\bf Example 1:}\hbr
The mirror geometry of $\IC^2/\IZ_n$:
Here $\cv$ is $\IP^1$ and the polynomial is given by
\eqn\ctwodual{
a_0x_1^n+a_1x_1^{n-1}x_2+\cdots +a_nx_2^n , }
so the hypersurface $\tilde X$ is just a collection of $n$ points in
$\IP^1$. 
A `singularity' of $\tilde X$ occurs whenever two or more of these points
coincide.
The secondary fan is determined by the columns of \lctwo.
For $n\ge 3$ we have to blow up the moduli space in order to have
a coordinate patch described by the large complex structure variables
$z_i=a_{i-1}a_{i+1}/(a_i)^2$ (with $1\le i\le n-1$).
The GKZ operators corresponding to the Mori cone generators,
\eqn\gkzactwo{\partial_{a_0}\partial_{a_2}-\partial_{a_1}^2,~~~
\partial_{a_1}\partial_{a_3}-\partial_{a_2}^2,~~\ldots~~,
\partial_{a_{n-2}}\partial_{a_n}-\partial_{a_{n-1}}^2}
become
\eqn\gkzzctwo{\Theta_{a_0}\Theta_{a_2}-z_1(\Theta_{a_1}-1)\Theta_{a_1},~~~
\ldots,~~~
\Theta_{a_{n-2}}\Theta_{a_n}-z_{n-1}(\Theta_{a_{n-1}}-1)\Theta_{a_{n-1}}}
with 
\eqn\thetactwo{\eqalign{&\Theta_{a_0}=\Theta_{z_1},~~~
    \Theta_{a_1}=-2\Theta_{z_1}+\Theta_{z_2}, \cr
    &\Theta_{a_i}=\Theta_{z_{i-1}}-2\Theta_{z_i}+\Theta_{z_{i+1}}
     ~~\hbox{for}~~2\le i\le n-3,\cr
    &\Theta_{a_{n-1}}=\Theta_{z_{n-2}}-2\Theta_{z_{n-1}},~~~
 \Theta_{a_n}=\Theta_{z_{n-1}}.}}
We note that the space of solutions of \gkzzctwo\ is too large unless
we introduce further operators corresponding to linear combinations of
the Mori cone generators.

The case of $n=2$ allows for an explicit solution \agmd : 
Here we have 
\eqn\ctwoztwogkz{{\cal D} = (\Theta_z-2z(2\Theta_z+1))\Theta_z}
and ${\cal D} = 0$ has a basis of solutions of the form
\eqn\ctwoztwosol{\varpi_0=1,~~~~
\varpi_1={1\over 2\pi i}\ln{1-\sqrt{1-4z}\over 1+\sqrt{1-4z}}.}
Special points in the moduli space are the large complex structure
limit $z=0$, the analog of the primary component of the discriminant
locus at $z=1/4$, and the orbifold point at $z=\infty$ where we
introduce a new coordinate $\varphi$ by $z\varphi^2=1$.
We find the following transformation properties upon taking loops
around these points:
\eqn\ctwoztwomon{
z=0:~~\varpi_1\to\varpi_1+1,~~~~
z=1/4:~~\varpi_1\to - \varpi_1,~~~~
\varphi\to e^{\pi i}\varphi:~~\varpi_1\to -1 - \varpi_1.}

\noindent {\bf Example 2:}\hbr
The local mirror geometry $\tilde X$ of the resolution $X$ of
$\IC^n/\IZ_n$ is just the mirror geometry of a compact Calabi-Yau
manifold realised as a degree $n$ hypersurface in $\IP^{n-1}$,
i.e. $\tilde X$ is a degree $n$ hypersurface 
\eqn\znmirror{a_1x_1^n+\cdots +a_nx_n^n +a_{n+1}x_1\ldots x_n=0}
in $\IP^{n-1}/(\IZ_n)^{n-2}$.
The GKZ operator
$\partial_{a_1}\ldots\partial_{a_n}-\partial_{a_{n+1}}^n$
becomes 
\eqn\zngkz{\Theta_z^n-z (-n\Theta_z-n+1)(-n\Theta_z-n+2)\cdots(-n\Theta_z)}
in terms of the large complex structure variable 
$z=a_1\ldots a_n/(a_{n+1})^n$.

\noindent {\bf Example 3:}\hbr
The mirror geometry of $\IC^3/\IZ_5$, a genus two Riemann surface:
\ifig\ptwozfive{The fan for $\IP^2/\IZ_5$.}
{\epsfxsize1.5in\epsfbox{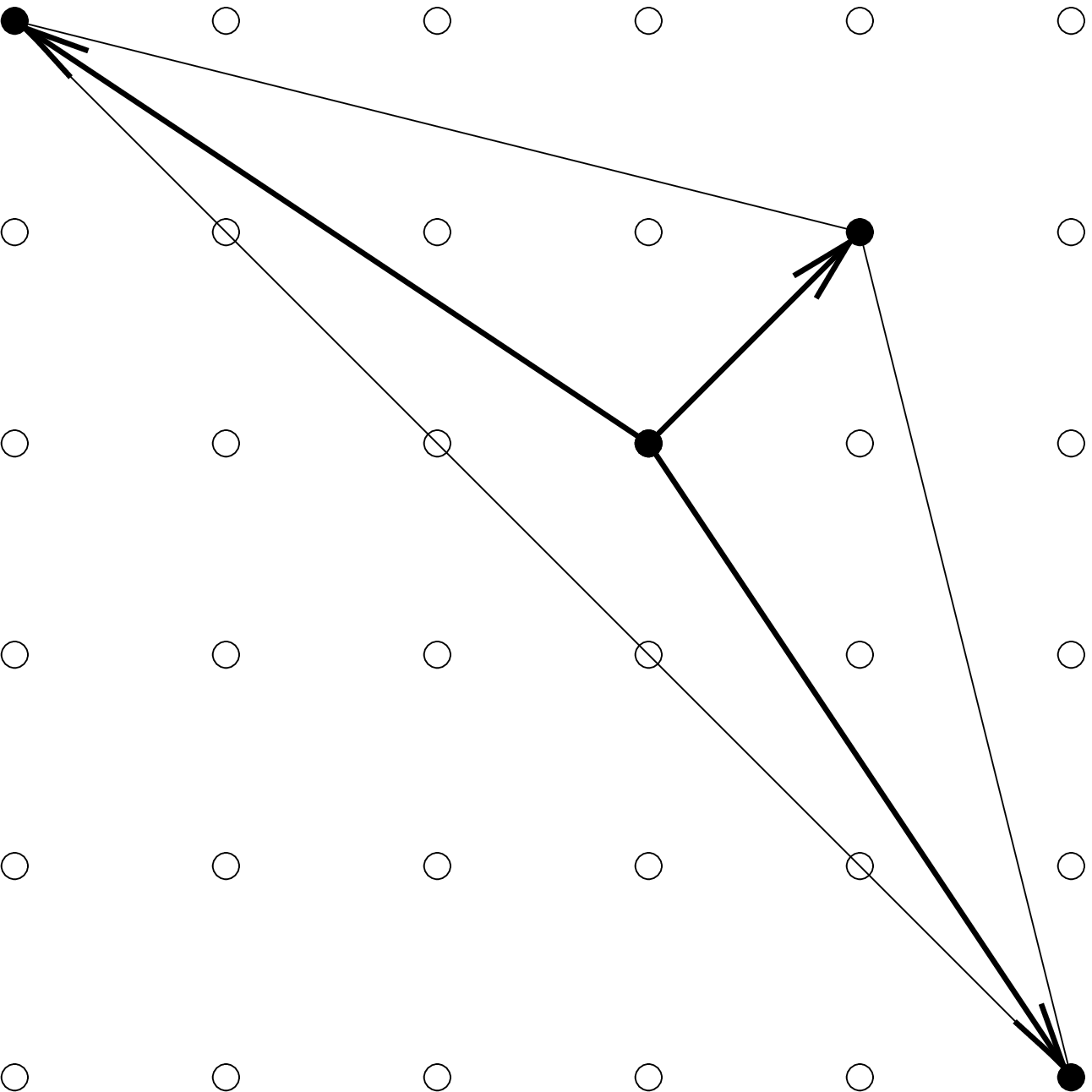}}
Here $\cv$ is $\IP^2/\IZ_5$ with the $\IZ_5$ acting on the homogeneous
coordinates of $\IP^2$ as $(x_1,x_2,x_3)\to (\ep x_1,x_2,\ep^{-1}x_3)$.
The polynomial corresponding to \fan\ is given by
\eqn\quin{
a_1x_1^5+a_2x_2^5+a_3x_3^5+a_4x_1^2x_2x_3^2+a_5x_1x_2^3x_3, }
where we have chosen the subscripts of the $a_i$ to correspond to
those of the $v_i$ in \fan .
The local mirror of $\IC^3/\IZ_5$ is given by the vanishing locus of
\quin\ in $\IP^2/\IZ_5$.
The action of $\IZ_5$ on $\IP^2$ has fixed points whenever two of the
three $x_i$ vanish. The vanishing locus of \quin\ passes through one
of these fixed points if and only if one of $a_1$, $a_2$, $a_3$
vanishes.
Thus the generic hypersurface misses the fixed points.
A quintic polynomial in $\IP^2$ defines, by a standard calculation, a
Riemann surface of Euler number $\chi=-10$. 
As the $\IZ_5$ acts without fixed points on this surface, the Euler
number is divided by 5, showing that the local mirror geometry is that
of a Riemann surface $\cal R$ with $2-2g=\chi=-2$, i.e. genus $g=2$.

Scalings $x_i\to \l_i x_i$ imply the equivalences
\eqn\modsp{(a_1,a_2,a_3,a_4,a_5)\sim 
(\l_1^5a_1,\l_2^5a_2,\l_3^5a_3,\l_1^2\l_2\l_3^2a_4,\l_1\l_2^3\l_3a_5).}
If we naively interpret the exponents of the $\l_i$ as a charge matrix 
\eqn\charges{
\left( \matrix{ 
5 & 0 & 0 & 2 & 1  \cr
0 & 5 & 0 & 1 & 3  \cr
0 & 0 & 5 & 2 & 1  \cr } \right) }
with entries $q_i^{j}$ and try to find a fan with rays $v_i$ fulfilling
$\sum_i q_i^{j}v_i=0$ for $j=1,2,3$, we find that we have to take
$v_1=v_3$, meaning that we should not distinguish between $a_1$ and
$a_3$.
This can be explained by the fact that taking $\l_3=\l_1^{-1}$ implies
that we can multiply $a_1$ with any nonzero number provided we divide
$a_3$ by the same number without affecting the other $a_i$, i.e. as
long as $a_1$ and $a_3$ are nonzero the complex structure of $\cal R$
depends only on $a_{13}:=a_1a_3$.
This is even true if one of $a_1,a_3$ becomes zero, since an exchange
of $a_1$ and $a_3$ can be compensated by exchanging $x_1$ with $x_3$
which does not affect the complex structure.
Thus we can consistently drop the third line and the third column of
\charges\ to obtain a matrix
\eqn\newcharges{
\left( \matrix{ 
5 & 0 & 2 & 1  \cr
0 & 5 & 1 & 3  \cr } \right). }
This is just the matrix of linear relations for the secondary fan of 
\secfan .
The corresponding compact toric variety 
\eqn\mtoric{
\cm_{\rm toric}~~=~~\left(\{(a_{13},a_2,a_4,a_5)\}\setminus
      \{(a_{13}=a_4=0)\vee(a_2=a_5=0)\}\right)~/~\sim}
with $\sim$ as in \modsp\ is closely
related to the moduli space $\cm_{\rm smooth}$ of smooth hypersurfaces
of the type \quin:
Smoothness implies $a_1\ne 0$, $a_2\ne 0$ and $a_3\ne 0$, so
\eqn\msmooth{
\cm_{\rm smooth}~~\subset~~\left(\{(a_{13},a_2,a_4,a_5)\}\setminus
      \{(a_{13}=0)\vee(a_2=0)\}\right)~/~\sim~~\subset~~ \cm_{\rm toric},}
i.e. $\cm_{\rm toric}$ is a compactification of $\cm_{\rm smooth}$
(other sensible compactifications correspond to omitting $v_{13}$ or $v_2$
from \secfan).
$\cm_{\rm smooth}$ is contained in the single coordinate patch of
$\cm_{\rm toric}$ defined by the cone spanned by $v_4$ and $v_5$.
In this patch we can parametrize the hypersurface as the vanishing
locus of 
\eqn\lmirror{
P_{\psi,\phi}(x_1,x_2,x_3)=
x_1^5+x_2^5+x_3^5-5\psi x_1^2x_2x_3^2 -5\phi x_1x_2^3x_3.}
Having set $a_1$, $a_2$ and $a_3$ to one has used up most of the
freedom coming from \modsp, the remaining relation being
\eqn\psphsim{(\psi,\phi) \sim (\ep^2\psi,\ep\phi).}
As we just noticed, $\cal R$ becomes singular along the divisors
$a_{13}=0$ and $a_2=0$ of $\cm_{\rm toric}$.
The remaining singularities can be found by looking for values of
$\psi$, $\phi$ where $\partial P_{\psi,\phi}/\partial x_i=0$ for
$i=1,2,3$ can be solved by some $(x_1,x_2,x_3)\ne(0,0,0)$.
This results in the equation
\eqn\conif{
16\psi^5+40\psi^4\phi^2+25\psi^3\phi^4+20\psi^2\phi+45\psi\phi^3+27\phi^5=1}
for the primary component of the discriminant locus.

We note that while $\cm_{\rm toric}$ contains some of the singular
loci, it misses others such as $a_1=a_4=0$, $a_2=a_5=0$ and any points
with three or four of the $a_i$ vanishing.
The divisor $a_{13}=0$ in $\cm_{\rm toric}$ corresponds to two one
dimensional loci $a_1=a_3=0$ and $a_1a_3=0$, $a_1+a_3\ne 0$.
Our main concern with the moduli space has to do with the study of
monodromies. 
Thus we want to know what happens when we move around singularities at
codimension one rather than what happens when we hit them.
For example, the monodromy around $a_{13}=0$ depends only on nonvanishing 
values of $a_{13}$ and not on how we interpret the locus $a_{13}=0$.
Therefore $\cm_{\rm toric}$ is sufficient for our purposes.

\ifig\disc{The moduli space of the resolution of $\IC^3/\IZ_5$.}
{\epsfxsize3in\epsfbox{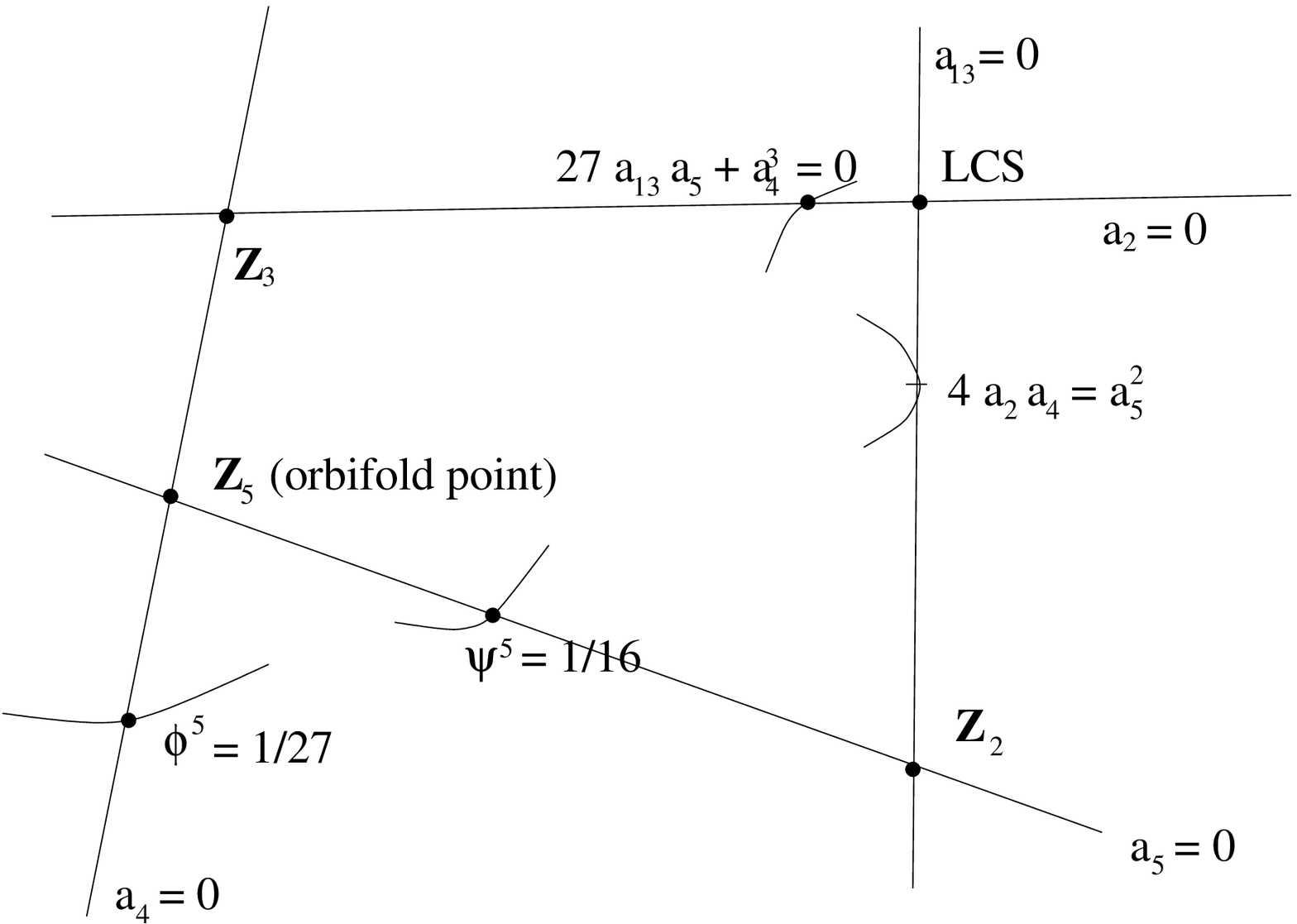}}

A schematic representation of $\cm_{\rm toric}$ is given in
\disc , with the the toric divisors shown as straight lines and the
primary component of the discriminant locus indicated by curved lines.
The locus $a_{13}=0$ is tangent to the discriminant locus at their point
of intersection $4a_2a_4=a_5^2$ whereas $27a_{13}a_5+a_4^3=0$
corresponds to a transverse intersection with $a_2=0$.
At $a_4=a_5=0$ (i.e., $\phi=\psi=0$) the moduli space has a
singularity where the Riemann surface remains smooth; in addition
there are $\IZ_2$ and $\IZ_3$ singularities at $a_{13}=a_5=0$ and
$a_2=a_4=0$, respectively. 

As we saw above, there is a distinguished
coordinate patch in $\cm_{\rm toric}$ which contains all loci where
$\cal R$ is smooth.
Now we want to study another distinguished set of coordinates
corresponding to the `large complex structure limit'.
We remember that the K\"ahler cone of $X$ (the resolution of
$\IC^3/\IZ_5$) was spanned by $D_1\sim D_3$ and $D_2$ corresponding to
the vectors $v_{13}$ and $v_2$ in the secondary fan (\secfan)
respectively.
The large radius limit of $X$ corresponds to the deep interior of the
K\"ahler cone, so by local mirror symmetry the large complex structure
limit is determined by the $v_{13}-v_2-$coordinate patch in 
$\cm_{\rm toric}$ given by
\eqn\ezes{
z_1=a_{13}{a_5\over a_4^3}={\phi\over 5^2\psi^3},~~~~
z_2=a_2{a_4\over a_5^2}=-{\psi\over 5\phi^2}.}
%Equivalently, we have
%\eqn\samezes{z_1={\phi\over 5^2\psi^3},~~ z_2=-{\psi\over 5\phi^2}.}
In terms of $z_1$, $z_2$ the principal component of the discriminant
locus is determined by 
\eqn\disczoneztwo{1 + 27z_1 - 8z_2 -   225z_1z_2 +   16z_2^2 + 
  500z_1{z_2}^2 +   3125z_1^2z_2^3=0.}
The GKZ system can be determined and solved with the methods described
above.
There are five independent solutions, as expected, which are described
in appendix A.

\newsec{D-branes and tautological bundles}
We want to find out about the D-brane vacuum states in type II string
theory on $X$.
The mathematical structure that captures the largest number of
properties of brane states is, at present knowledge, the bounded derived
category $D^b$ of coherent sheaves on $X$ \categ\AL\ (but see the
remarks in \CIL\DED ).
While we will make several remarks concerning $D^b$, we will work
mainly with the somewhat coarser (but easier to handle) concepts of
K-theory.
Let $K(X)$ be the Grothendieck group of coherent sheaves on $X$. 
We expect compact brane states on a non-compact space $X$ to correspond to
classes of the compactly supported K-theory group $K^c(X)$. 
Using the duality between $K(X)$ and $K^c(X)$ we can determine
a basis for $K^c(X)$ by first finding a basis for $K(X)$.

Let us consider the situation where $X$ is a smooth crepant resolution
of a singularity of the type $\IC^d/G$, where $G$ is a
finite subgroup of $SU(d)$.
Since $X$ is smooth, $K(X)$ is generated by vector bundles (see
e.g. \fultoni ).  
Moreover, if $\pi :X\rightarrow \IC^d/G$ is a crepant
resolution of an abelian singularity, $K(X)$ is in fact generated by $n$
line bundles, where $n$ is the order of $G$ 
(at least for $d\le 3$) \itonakajima .  
Thus, for finding a basis for the group $K^c(X)$ related to fractional
branes it is convenient to first determine a set $\{R_i\}$ 
($0\le i<n$) of line bundles whose K-theory classes generate $K(X)$.
Clearly there is no choice for the $R_i$ that should be preferred a
priori.
Rather, there are two distinct constructions, each of which is
related to McKay correspondence:
\smallskip
\noindent
{\it 1. Mathematicians' construction} \gsv--\itonakajima:
There is a vector bundle $R$ (the `tautological vector bundle')
transforming in the regular representation of $G$ whose
decomposition into irreducibles gives the line bundles $R^M_i$.
In particular, the $R^M_i$ are generated by their sections and the
action of $G$ on the sections determines a one-to-one correspondence 
between the $R^M_i$ and the characters of the irreducible
representations of $G$. 
In the case of a resolution of $\IC^2/G$ with some finite group
$G$ the first Chern classes $c_1(R^M_i)$, $i\ge 1$ form a basis of
$H^2(X,\IZ)$ dual to the basis of $H_2(X,\IZ)$ given by the homology
classes of a basis of effective curves $C_i$ in the resolution. 
In the case of a singularity of the type $\IC^3/G$ with $G$
an abelian subgroup of ${\rm SL}(3,\IC )$ in general there exist
several crepant resolutions and not for every resolution it is
possible to define line bundles as above. However, it was shown in 
\nakamura\ that there exists a distinguished crepant resolution, 
named $G$-Hilb, on which it is still possible to define the
tautological line bundles (see also \reidmk,\crawreid,\craw)\foot{We
thank A. Craw for emphasizing the importance of choosing the G-Hilb
resolution to us.}. 
The advantage of this approach is that it is rigorously proven for
$d=2$ and $d=3$. 

\smallskip
\noindent
{\it 2. Physicists' constructions}:
The authors of \ddd\ suggest to consider, in the style of \dgm, the
world-volume theory of D0-branes, which is a theory of 
$n-1$ $U(1)$ gauge fields and $d$ $n\times n$ matrices.
It is conjectured (and shown in several examples) that the vacua of
 such a theory in the different phases corresponding to different
choices of Fayet-Iliopoulos parameters all lead to moduli spaces that
are nothing but the geometric phases of the resolutions $X$ of
$\IC^d/G$.
Now repeat this construction with an extra field transforming in a
specific one dimensional representation $\rho_i$ of $G$.
It is conjectured that, independently of the phase, this should lead to
a space that is the total space of a line bundle $R_i^P$ over $X$,
and that repeating this for all characters $\rho_i$ should give a
basis $\{R_i^P\}$ of $K(X)$. However, this construction is extremely
tedious to work with.  
%\par 
A different method for determining $\{R_i^P\}$ based on the
boundary chiral ring associated to a certain two dimensional gauge
theory has been proposed in \mayr . The implications of this approach
have been worked out for the case of a single exceptional divisor that 
is a weighted projective space $W=\IP^{d-1}_{n_1,\ldots,n_d}$ with Fermat 
weights \mayr\ or a Grassmannian\foot{P. Mayr informs us that this
approach works in more general situations as well.}\LMW . 
\del
In the former particular situation the physicists'
approaches lead to bundles whose restrictions to $W$ are given
by $\co$, $\co(-1),\ldots,\co(-n+1)$, with $n=\sum n_i$. 
\enddel
In all examples we are aware of, the $R_i^P$ have no sections.
The advantage of this approach is that it appears to lead to dual
classes $S_i^P$ whose interpretations in terms of D-branes are very
well behaved. 

\smallskip

Roughly, the resulting $R_i$ can be summarized in the following way.
There is a set of divisor classes $\{[F_i]\}$ containing all K\"ahler
cone generators $[T_i]$ and the trivial class $[0]$ such that all
$F_i$ are nef, i.e. have nonnegative intersection with any curve in
the Mori cone. 
If we denote by $R_i^\pm$ the line bundles $\co(\pm F_i)$, then 
$\{R_i^M\}=\{R_i^+\}$ and $\{R_i^P\}=\{R_i^-\}$.
In two dimensions the $[F_i]$ are just the trivial class and the
K\"ahler cone generators.
In higher dimensions we have to add extra divisor classes which are
nonnegative integer linear combinations of the $[T_i]$.
For the $R_i^M$ with G-Hilb and $d=3$ the authors of
\reidmk,\crawreid\ have given an explicit construction. 
In terms of the language used in this paper this can be summarised in
the following way.

Through the sections we can assign a character to any $T_i$.
It is also possible to assign characters to toric curves.
Such a curve $C$ corresponds up to a sign to some $m\in M$ leading to a
linear equivalence as in \lineq.
By collecting expressions with the same sign this can be written as
$D\sim D'$ where $D$, $D'$ are effective divisors corresponding to the
same character. 
We then assign this character to $C$ and the corresponding line
segment in the diagram, and find that all the characters
obtainable in this way also occur in the list of characters
corresponding to the $T_i$.
Then every interior point $I$ of the toric diagram is of one of the
following types:\hbr
1. There are three pairs of line segments with the same character meeting 
in $I$.
In this case we add nothing to the list of $[F_i]$ (the classes assigned
by \reidmk,\crawreid\ in this case are already among the K\"ahler cone 
generators).\hbr
2. There are two pairs of line segments with characters $\chi_m$,
$\chi_n$ meeting in $I$ (and possibly an extra line segment). 
Then add $[T_m+T_n]$  to the list of $[F_i]$.\hbr
3. There are three line segments with the same character $\chi_m$.
In this case add $[2T_m]$  to the list of $[F_i]$.\hbr
It turns out that this procedure always leads to a one to one
correspondence between the $R_i^+$ and the character table of $G$
through the action of $G$ on the sections.

In many cases the $[F_i]$ are the same in the mathematicians' and
physicists' constructions, i.e. $R_i^P=(R_i^M)^*$.
However, \mayr\ seems to suggest partial resolutions in the case with
a single interior point where the exceptional divisor is a weighted
projective space. 
We note that the G-Hilb resolution may be incompatible with such a
resolution or any refinement of it, as the following example shows.
\ifig\pott{$G$-Hilb and partial resolution of $\IC^3/\IZ_6$.}
{\epsfxsize2.5in\epsfbox{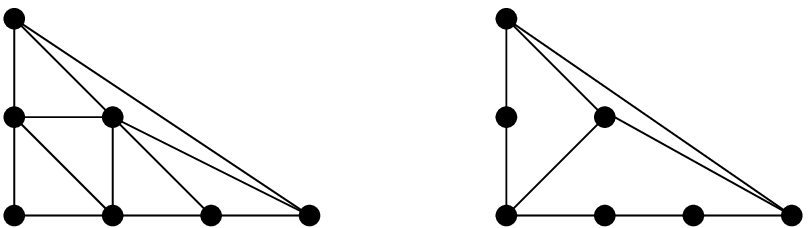}}
In \pott\ we have displayed the G-Hilb resolution of $\IC^3/\IZ_6$
constructed according to the rules of \reidmk, \crawreid\
and the partial resolution by an exceptional divisor
$\IP^2_{(1,2,3)}$.
Clearly the former cannot be obtained as a refinement of the latter.

In the following we always follow the mathematicians' approach.

\del
Then we define a set $\{[F_i]\}$ of divisor classes by taking the
union of the set of the K\"ahler cone generators $\{T_i\}$ and the set
$\{D_m\}$ of
divisor classes associated to interior points as above. 
Then we define collections $\{R^+\}$ and $\{R^-\}$ of line bundles by
$R^\pm_i=\co_X(\pm F_i)$.
\smallskip
We observe the following facts:\hbr
a) In all examples we are aware of, the $R_i^+$ are the same as the
$R_i^M$.\hbr
b) In all examples we are aware of, the $R_i^-$ are the same as the
$R_i^P$ (admittedly, $R_i^P$ isn't known in too many cases). In
particular, our procedure reduces to that of \mayr\ in the case of a
single compact exceptional divisor which is a weighted projective space with
Fermat weights. \hbr 
c) There is always a one to one correspondence between the
$R_i^+$ and the character table of $G$ through the action of
$G$ on the sections.\hbr
\enddel

The next step in our construction of D-brane states is to find a basis
for $K^c(X)$ that is dual to the basis of $K(X)$ defined in terms of
line bundles $R_i$.
According to \itonakajima , there is a pairing $(R,S)$
between representatives $R$ of $K(X)$ and $S$ of $K^c(X)$ that can be
evaluated in terms of Chern characters
\eqn\itonaka{
(R,S)=\int_X {\rm ch}(R)\cup{\rm ch}^c(S)~{\rm Td}(X),
}
with ${\rm ch}^c(S)$ the localized Chern character\foot{
Let $i:Y\hookrightarrow X$ be the embedding of a compact submanifold
$Y$ in a noncompact manifold $X$. Elements of 
the compactly supported K-theory can be represented by either coherent 
sheaves $S_Y$ on $Y$ or by their finite resolution by vector
bundles on $X$, that is by complexes $S$ of vector bundles on $X$
which are exact off $Y$ and whose homology is precisely the
push-forward of $S_Y$ to $X$ \bfmp . Then, the local Chern
character is defined such that ${\rm ch}^c(S)={\rm ch}(i_*S_Y)$
\iversen .}  
of the complex $S$ and 
${\rm Td}(X)$ the Todd class of $X$.
There is also a closely related pairing which will become important
when we study monodromies. It is defined as
\eqn\skewpair{\< R,S \>=(R^*,S)}
with $R^*$ the line bundle (or, more generally, the complex) dual to $R$.
If we restrict $R$ to $K^c(X)$, these pairings become well defined
under the exchange of $R$ and $S$ and we find that $(R,S)$ is always
symmetric whereas $\< R,S \>$ is symmetric in even dimensions and skew
in odd dimensions, as a consequence of the fact that ${\rm Td}(X)$ is
even when $c_1(X)$ is trivial.

The generally accepted way of obtaining a basis for $K^c(X)$ is to choose
classes dual to those given by the line bundles $R_i$ with respect to
$(~~,~~)$.
Following this convention, we define classes of $K^c(X)$ by demanding
that their representatives $S_j$ fulfill $(R_i,S_j)=\delta_{ij}$.
Thus we obtain $S_j^+$ dual to $R_i^+$ and $S_j^-$ dual to $R_i^-$
with respect to $(~~,~~)$ and note that the $S_j^+$ are dual to the
$R_i^-$ and $S_j^-$ are dual to the $R_i^+$ with respect to
$\<~~,~~\>$.

So far we have not been specific about the representatives $S_i$ of
the compactly supported K-theory.
In the spirit of \DK\ we may interpret them as bound states of
$X$--filling branes.
In mathematical terms this amounts to specifying a complex of vector
bundles on $X$ that is exact outside a compact locus $Y$.
It is not hard to check in every example that we may indeed represent 
every $S_i$ as a formal linear combination of line bundles of the form 
$\co_X(\sum a_i D_i)$ and
that the ${\rm ch}^c(S_i)$ obtained from the line bundles $R_i$ form a
basis for all Chern characters with support on the compact toric cycles.

Alternatively, one may wish to consider `pure' branes defined in terms
of the structure sheaves of the independent lower dimensional compact
holomorphic cycles. 
Given the structure sheaves $\co_{\cc_i}$ where the $\cc_i$ form a basis
for all compact holomorphic cycles on $X$, applying $k$ push-forwards
for every cycle of codimension $k$ leads to sheaves $\tilde S_{\cc_i}$ on
$X$.
In order to relate these objects to Chern characters on $X$ we have to
use the Grothendieck-Riemann-Roch theorem, 
\eqn\GRR{i_*({\rm ch}(S_D){\rm Td}(D))={\rm ch}(i_*S_D){\rm Td}(X)}
for embeddings of the type $i:D\hookrightarrow X$.
Writing ${\rm ch}^c(S)=\sum n^{(\cc_i)} {\rm ch}(\tilde S_{\cc_i})$ 
allows us to define the charge vectors $\vec n(S)$.
Alternatively we may calculate the charge vectors by first calculating
\eqn\rstilde{
(R_j,\tilde S_{\cc_i})=
\int_X {\rm ch}(R_j){\rm ch}(\tilde S_{\cc_i}){\rm Td}(X)=
\int_{\cc_i} {\rm ch}(R_j|_{\cc_i}){\rm Td}(\cc_i)=
\chi(R_j|_{\cc_i},\cc_i)=:\chi_{ji}}
and noticing that $(R_j,S_k)=\delta_{jk}$ implies 
$\sum_i\chi_{ji}n^{(\cc_i)}_k=\delta_{jk}$, i.e. 
$n^{(\cc_i)}_k=(\chi^{-1})_{ik}$.
We note that the compact holomorphic cycles generate the
compact homology of $X$, so the number of $\cc_i$ is equal to
$\chi^c(X)=\sum_{i=0}^{2d-1}(-1)^iH^c_i$ which is just the number of
$d$-dimensional cones in $\Sigma$ \ful.

At the large volume limit the Mukai vector ${\rm ch}^c(S)\sqrt{{\rm Td}(X)}$
determines the central charge\foot{This formula occurs implicitly in
\dr\ and explicitly in \ddd; see also the remarks in \mayr.} 
\eqn\ccharge{
Z^{\rm lv}(t_i;S)=-\int_Xe^{-\sum t_i T_i}{\rm ch}^c(S)\sqrt{{\rm Td}(X)}}
of the brane configuration, where the $T_i$ are generators of the
K\"ahler cone. 
In particular we obtain 
\eqn\cchargepc{
Z^{\rm lv}(t;\tilde S_p)=-1~~~\hbox{and}~~~Z^{\rm lv}(t;\tilde S_{C_i})=t_i-1}
for the central charges of D0-branes and D2-branes wrapping (with
trivial bundle) the generators $C_i$ of the Mori cone dual to the $T_i$.
These are the objects related by local mirror symmetry to the
solutions of the GKZ system at the large complex structure point.
More precisely we expect the exact central charge $Z(z;S)$ to be a
linear combination of the GKZ solutions such that 
\eqn\lcslim{\lim_{z\to 0}(Z(z;S)-Z^{\rm lv}(t_i;S))=0.}
If we demand that $Z^{\rm lv}(t;\tilde S_{C_i})$ measure the complexified
K\"ahler class at the large K\"ahler limit we have to make the
identification 
\eqn\ident{t_i-1={\ln z_i\over 2\pi i}+O(z).}
Note that this is different from the conventions usually adopted
in the literature, but we find that this is precisely the
identification that works.

Linearity implies that the central charge corresponding
to any $S$ is given in terms of the charge vector by
\eqn\ZQ{Z(S)=\sum n^{(\cc_i)}  Z(\tilde S_{\cc_i}).}

Finally, we return to the subject of monodromy.
In \konts\ it was conjectured (and pushed further in the work of \paul\psrpt ) 
that the monodromies around loci in
the moduli space where the mirror $\tilde X$ of a Calabi-Yau threefold
$X$ becomes singular induce autoequivalences of $D^{b}(X)$, the
bounded derived category of coherent sheaves on $X$. 
Moreover, in the case of a Fano surface
embedded in a Calabi-Yau threefold, a relationship of these
autoequivalences of $D^{b}(X)$ with mutations of exceptional
collections supported on the Fano surface was pointed out in \psrpt .
For our purposes we will view the various monodromies mainly as
automorphisms of $K^c(X)$. However, in some examples we will identify
the monodromy actions on the exceptional collections of coherent
sheaves supported on the compact divisors. 
As in the case of the local mirror geometry, we will be interested in
the following three types of transformations:\hbr
--- Monodromy around large K\"ahler structure divisors in the moduli 
    space,\hbr
--- Monodromy around the primary component of the discriminant
    locus,\hbr
--- `Orbifold monodromy' in the case $\IC^d/G$.

Only the monodromy around a divisor $z_i=0$ in the moduli space where
the K\"ahler parameter $t_i$ (associated with the divisor class $[T_i]$
in $X$) becomes infinite allows for a classical analysis.
In this case we just take $t_i\to t_i+1$ in \ccharge. 
Because of the multiplicativity of Chern characters, the fact that the
Chern character of a line bundle is the exponential of its first Chern
class and the form of \ccharge , this transforms the $S_j$ by tensoring
them with $\co_X(-T_i)$.
By \itonaka , the $R_i$ transform by tensoring with $\co_X(T_i)$.

According to the observations in \aspi\dg , `orbifold monodromy' should
cyclically permute the $S_i$ if $X$ is a resolution of $\IC^d/\IZ_n$.

For the primary component of the discriminant locus we have the following
picture: 
In the case of a compact Calabi-Yau variety $X$ it is conjectured
(see \konts\paul\psrpt\thomass\aspmon) 
that a sheaf $\cal F$ is subjected to a Fourier-Mukai transform whose
kernel is the structure sheaf $\co_X$, implying that the Chern character
of $\cal F$ transforms as 
\eqn\foumukco{
{\rm ch}({\cal F})\to{\rm ch}({\cal F})-\<\co_X, {\cal F}\>{\rm ch}(\co_X),}
where $\<~~,~~\>$ is the pairing \skewpair.
In our case of non-compact $X$ this cannot work because it would
violate compact support conditions, but we make the following
observation:

{\it In all of our examples we obtain expressions for 
${\rm ch}^c(S_0^-)$ that allow us to choose $S_0^-$ in such a way that
its restriction $S_0^-|_{\cc_i}$ to any compact toric cycle $\cc_i$ is
equal to $\co_{\cc_i}$}.
For the case of a resolution $\pi$ of an orbifold singularity this
means that our expressions for ${\rm ch}^c(S_0^-)$ are consistent
with taking $S_0^-$ to be the push-forward of the restriction of
$\co_X$ to $\pi^{-1}(0)$.

{\it Wherever we have the possibility of comparison with the mirror
geometry, we find that the monodromy around the primary component of
the discriminant locus is given by} 
\eqn\foumuknc{
{\rm ch}({\cal F})\to{\rm ch}({\cal F})-\<S_0^-, {\cal F}\>{\rm ch}^c(S_0^-).}
More precisely, the following happens:
For one parameter models the principal component is pointlike.
If we decompose the GKZ solutions into logarithms and holomorphic
pieces at $z=0$, the principal component is at the boundary of the
radius of convergence of the holomorphic pieces.
In this case we find that the monodromy is given precisely by
\foumuknc\ provided we choose the simplest anti-clockwise path,
$\ln(-1)=\pi i$ and the identification \ident.
With more than one parameter the discriminant locus consists of
several disjoint pieces in the $z$ coordinate patch (these pieces join
in the other coordinate patches), and there is no unambiguous choice of 
component or path.
We find, however, that at every branch one of the $S_i^-$ (possibly
transformed by large complex structure monodromy) becomes massless.
This is consistent with the picture that when we take $S_0^-$ along
some non-trivial paths like the ones corresponding to `orbifold
monodromy' we turn it into one of the other $S_i^-$.

If $d$ is even there is a simple consistency check: If we require
\foumuknc\ to respect the pairing $\<~~,~~\>$ then it is easily checked
that this is equivalent to $\<S_i^-,S_i^-\>=2$ (in odd dimensions the
analogous condition $\<S_i^-,S_i^-\>=0$ is fulfilled automatically
because of the skew symmetry of $\<~~,~~\>$).
This is indeed true in all of our examples.

\noindent {\bf Example 1:}\hbr
Resolution of $\IC^2/\IZ_n$:
The case of $\IC^2/G$ with $G$ any discrete subgroup of
$SU(2)$ is well understood by mathematicians in the context of McKay
correspondence.
If $G$ is abelian and resolved by the introduction of a set
$\{C_i\}$ of exceptional curves, and if $\{[C_i^\vee]\}$ is a
basis of divisor classes dual to $\{C_i\}$ then the $R^M_i$ are given
by $R^M_0=\co_X$ and $R^M_i=\co_X(C_i^\vee)$ for $i\ge 1$. 
By \kahlctwo\ sections of the $R_i^+$ are given, for example, by
$1,~z_0,~z_0^2z_1,~z_0^3z_1^2z_2,~\ldots$, so the action of $\IZ^n$
on these sections through $z_0\to \ep z_0$, $z_n\to \ep^{n-1}z_n$ indeed
reproduces the characters $1,~\ep,~\ep^2,~\ldots$ of $\IZ^n$.

Using \itonaka\ and denoting by $p$ the class of a point, we find 
\eqn\ctwoS{
{\rm ch}^c(S_0^\pm)=p\mp\sum_{i=1}^{n-1}C_i,~~~
{\rm ch}^c(S_i^\pm)=\pm C_i~~\hbox{for}~~i>0,~~~~~
{\rm ch}(\tilde S_{C_i})=p+C_i,~~~ {\rm ch}(\tilde S_p)=p}
and therefore
${\rm ch}^c(S_0^-)=
\sum_{i=1}^{n-1}{\rm ch}(\tilde S_{\cc_i})-(n-2){\rm ch}(\tilde S_p)$.
The restriction of $\co_X$ to the union of the $C_i$ is the same as
$\sum\tilde S_{C_i}$ except for the $n-2$ points of the form 
$C_i\cdot C_{i+1}$ where $\sum\tilde S_{C_i}$ has rank two.
Upon subtracting the $n-2$ sheaves with support on these points we
arrive at a class that matches ${\rm ch}^c(S_0^-)$.
It is easily checked that $\<S_i^-,S_i^-\>=2$ for all $i$.
The large volume central charges are given by 
$Z^{\rm lv}(t;S_0^-)=-1+\sum_i t_i$ and $Z^{\rm lv}(t;S_i^-)=-t_i$.

In the case of $n=2$ this implies  $Z(S_0^-)=\varpi_1$ and 
$Z(S_1^-)=-1-\varpi_1$ and we see that the principal component and orbifold
monodromies found in the mirror geometry are precisely the ones
generated by \foumuknc\ and permutations $S_0^-\leftrightarrow S_1^-$,
respectively.

\noindent {\bf Example 2:}\hbr
For $\IC^n/\IZ_n$ with $\IZ_n: (z_1,\ldots,z_n) \to 
(e^{2\pi i/n}z_1,\ldots,e^{2\pi i/n}z_n)$ the restrictions of the
$R^M_i$ to the exceptional divisor $D\simeq\IP^{n-1}$ are nothing but 
$\co$, $\co(1),\ldots,\co(n-1)$. 
The independent holomorphic cycles are of the form $\IP^j$ with 
$0\le j\le n-1$ and the $R_i^\pm$ restrict to $\co_{\IP^j}(\pm
i)$. This example has been previously considered in \zaslow\VHI\mayr
. We include it as further evidence that the $S_i$ have the properties
stated above. Defining 
\eqn\chikj{\chi_{kj}:=\chi(\co_{\IP^j}(k),\IP^j)=
\int_{\IP^j}{\rm ch}(\co_{\IP^j}(k)){\rm Td}(\IP^j)=
\int_{\IP^j}e^{kH}\left({H\over 1-e^{-H}}\right)^{j+1}}
with $H$ the hyperplane divisor, we find that 
\eqn\chikjev{\eqalign{\chi_{kj}-\chi_{k-1,j}=
&\int_{\IP^j}e^{kH}(1-e^{-H})\left({H\over
1-e^{-H}}\right)^{j+1}=\cr
&\int_{\IP^j}e^{kH}H\left({H\over
1-e^{-H}}\right)^j=
\int_{\IP^{j-1}}e^{kH}\left({H\over
1-e^{-H}}\right)^j=
\chi_{k,j-1}.}}
With $\chi_{k0}=\chi_{0j}=1$ this simple recursion is solved by
$\chi_{kj}=({k+j\atop j})$ and we obtain 
$(R_i^\pm,\tilde S_{\IP^j})=({j\pm i\atop j})$, implying
$(R_0^-,\tilde S_{\IP^j})=1$ for any $j$,  
$(R_i^-,\tilde S_{\IP^j})=0$ for $1\le i\le j$ and
$(R_i^-,\tilde S_{\IP^j})=(-1)^j({i-1\atop j})$ for $i>j$.
This leads to the following expressions for the $S^-$:
\eqn\smincd{S_0^-=\tilde S_{\IP^{n-1}}, ~~~~~
(-1)^kS_k^-=\left({n-1\atop k}\right)\tilde S_{\IP^{n-1}}-
\sum_{j=k-1}^{n-2}\left({j\atop k-1}\right)\tilde S_{\IP^j}
~~~\hbox{for}~~k\ge 1.}
Again the restriction of $S_0^-$ to any compact toric cycle is the
same as the structure sheaf of that cycle.

Alternatively we may determine the $S_i^-$ by the ansatz 
$S_i^-=\sum a_{ik}i_*\co_{\IP^{n-1}}(k)$.
With 
\eqn\riistar{
(R_i^-,i_*\co_{\IP^{n-1}}(k))=\chi_{k-i,n-1}=\left({n-1+k-i\atop n-1}\right)}
we get $(a^{-1})_{ki}=\chi_{k-i,n-1}$ which leads to
\eqn\cnsi{
S_i^-=\sum_{k=0}^n a_{ik}i_*\co_{\IP^{n-1}}(k)=
\sum_{k=0}^i (-1)^{i-k}\left({n\atop i-k}\right)i_*\co_{\IP^{n-1}}(k);}
%Now we can calculate 
\eqn\sisi{\eqalign{
\<S_i^-,S_i^-\>=&\sum_{k,l}a_{ik}~a_{il}\int_X {\rm ch}
(i_*\co_{\IP^{n-1}}(k))^* {\rm ch}(i_*\co_{\IP^{n-1}}(l)) {\rm Td}(X)=\cr
&\sum_{k,l}a_{ik}~a_{il}\int_{\IP^{n-1}}(1-e^{-nH})~e^{-kH}~e^{lH}
{\rm Td}(\IP^{n-1})=\cr
&\sum_{k,l}a_{ik}~a_{il}~(\chi_{l-k,n-1}-\chi_{l-k-n,n-1}).}}
Using $\chi_{l-k-n,n-1}=
%\left({l-k-1\atop n-1}\right)=
%(-1)^{n-1}\left({n-1+k-l\atop n-1}\right)=
(-1)^{n-1}\chi_{k-l,n-1}$ and the reciprocity of $a$ and $\chi$ we get 
\eqn\cnss{
\<S_i^-,S_i^-\>=a_{ii}(1-(-1)^{n-1})=0/2~~\hbox{ for $n$ odd/even,}}
as it should be.

For $n=3$ we find $Z^{\rm lv}(t;S_0^-)=-t^2/2+3t/2-5/4$.
The corresponding GKZ system has been studied at various places in the
literature, e.g. in \agmd\ and \dg .
In terms of solutions 
\eqn\gkzcthree{\varpi_0=1,~~~ \varpi_1={\ln z\over2\pi i} +O(z),~~~ 
\varpi_2=\left({\ln z \over2\pi i}\right)^2 +O(z\ln z),}
the rule $t\sim 1+\ln z/(2\pi i)$ leads to
$Z(S_0^-)=-\varpi_2/2+\varpi_1/2-1/4$.
Comparing with \dg, we find that this is precisely the expression
denoted there by $t_d$ which vanishes at the discriminant point $z=-1/27$.

\noindent {\bf Example 3:}\hbr
The K\"ahler cone is generated by $[D_1]$ and $[D_2]$ corresponding to the
characters $\ep=e^{2\pi i/5}$ and $\ep^3$, respectively.
By applying the rules outlined above, we assign the character $\ep$
to each of the three line segments meeting at $v_4$ in \fan\ and to the line
segment between $v_5$ and $v_2$, whereas the remaining two line segments (from
$v_5$ to $v_1$ and $v_3$) correspond to $\ep^3$.
Thus we get $F_2= 2D_1$ because of the three line segments with equal
characters meeting at $v_4$ and $D_1+D_2$ because of the two pairs of line
segments at $v_5$.
Altogether we get representatives $R_i^\pm=\co(\pm F_i)$ with
\eqn\thers{
F_0=0,~~F_1= D_1,~~F_2= 2D_1,~~F_3= D_2,~~F_4=D_1+D_2}
for the bases of $K(X)$, where we have chosen the labels such that
sections of $R_i^+$ transform as $\ep^i$ under 
$(z_1,z_2,z_3)\to(\ep z_1,\ep^3 z_2, \ep z_3)$.
Using \itonaka\ we find that the localized Chern
characters of the basis of $K^c(X)$ are given by
\eqn\sxheaves{\eqalign{
& {\rm ch}^c(S_0^\pm)=D_4+D_5\mp ({3\over 2}h+{5\over 2}f)+{11\over 6}p, \cr
& {\rm ch}^c(S_1^\pm)=-2D_4-D_5\pm (2h+{3\over 2}f)-{4\over 3}p, \cr
& {\rm ch}^c(S_2^\pm)=D_4\mp {1\over 2}h+{1\over 2}p, \cr
& {\rm ch}^c(S_3^\pm)=-D_5\pm {5\over 2}f-{1\over 3}p, \cr
& {\rm ch}^c(S_4^\pm)=D_5\mp {3\over 2}f+{1\over 3}p,
}}
with $p$ the class of a point.

Let us now consider the branes defined in terms
of the structure sheaves $\co_p$, $\co_h$, $\co_f$, 
$\co_{\IP^2}$, $\co_{\IF_3}$ of the independent lower dimensional
cycles.
Denoting by $\tilde S_p$ the result of three succesive
inclusion maps acting on $\co_p$, etc., we arrive with the help
of the Grothendieck-Riemann-Roch theorem \GRR\ at the
following result: 
\eqn\stsheaves{\eqalign{
& {\rm ch}(\tilde S_{D_4})=D_4+{3\over 2}h+{3\over 2}p, \cr
& {\rm ch}(\tilde S_{D_5})=D_5+h+{5\over 2}f+{4\over 3}p, \cr
& {\rm ch}(\tilde S_p)=p, \cr
& {\rm ch}(\tilde S_h)=h+p, \cr
& {\rm ch}(\tilde S_f)=f+p.
}}
This allows us to determine the D-brane charges 
$n_i=(n_{D_4}, n_{D_5}, n_p, n_h, n_f)$ with $n_p$ the
D0-brane charge, $n_h, n_f$ D2-brane charges and $n_{D_4}, n_{D_5}$ 
D4-brane charges of the $S_i^-$ as
\eqn\nch{\eqalign{
& n_0=(1, 1, 0, -1, 0) \cr
& n_1=(-2, -1, 0, 2, 1) \cr
& n_2=(1, 0, 0, -1, 0) \cr
& n_3=(0, -1, 0, 1, 0) \cr
& n_4=(0, 1, 1, -1, -1).
}}
In particular, this means that 
$S_0^-=\tilde S_{D_4}+\tilde S_{D_5}-\tilde S_h$.
Note how $\tilde S_{D_4}+\tilde S_{D_5}$ has rank 1 on $D_4$ and on
$D_5$ except on their intersection $h$, where it has rank 2 which is
compensated by subctracting $\tilde S_h$.

At this point we would like to mention that we have performed a
similar analysis for $\IC^3/\IZ_n$ with arbitrary odd $n$ and an
action of the type 
$(z_1,z_2,z_3)\to (\epsilon z_1,\epsilon^{n-2} z_2,\epsilon z_3)$.
In that case the resolution requires a $\IP^2$ and $(n-3)/2$
Hirzebruch surfaces and we again obtain $R_i^+$ whose sections
transform by the characters of $\IZ_n$ and an expression for $S_0^-$
that reduces to the structure sheaf on every compact toric cycle.
\del
The $n_i$ for the $S_i^+$ are more complicated, so
from now on we will only work with the $S_i^-$, dropping the
corresponding superscript.
\enddel

Returning to $\IC^3/\IZ_5$ we now give an alternative description of
the compactly supported K-theory classes in terms of non-trivial
sheaves on the exceptional divisors. 
Again with the help of the Grothendieck-Riemann-Roch theorem, we find
that we may choose representatives $S_i$
%for the classes of $K^c(X)$ dual to the classes represented by the $R_i$ 
in terms of the following combinations of push-forwards of sheaves:
\eqn\sdsheaves{\eqalign{
& S_0^-=j_*(\co_{D_5}(-h))+ g_*\co_{D_4}, \cr
& S_1^-=-j_*(\co_{D_5}(-h-f))-g_*V, \cr
& S_2^-=g_*\co_{D_4}(-1), \cr
& S_3^-=-j_*\co_{D_5}(-h), \cr
& S_4^-=j_*\co_{D_5}(-h-f),
}}
with $g:D_4\hookrightarrow X$ and $j:D_5\hookrightarrow X$ inclusion
maps. By $V$ we denote a stable bundle on $\IP^2$ with the Chern
character given by
\eqn\v{{\rm ch}(V)=2-h-{1\over 2}p}
where $p$ is the class of a point on $\IP^2$. 
Note that $\{\co_{\IP^2}(-1), V, \co_{\IP^2}\}$ is a foundation of the
helix of exceptional bundles on $\IP^2$ and that 
$\{\co_{\IF_3}(-h-f), \co_{\IF_3}(-h)\}$ is a regular exceptional pair
on $\IF_3$ (see \kvino ).

\del
We can now compute the Mukai vector charge \mukaiv\
and the central charge of the brane configuration that corresponds to
the above charge vector
\eqn\ccharg{
Z^{\rm lv}(t_1,t_2;S)=-\int_Xe^{-(t_1D_1+t_2D_2)}{\rm ch}(i_*S)\sqrt{{\rm Td}(X)}~.
}
Note that \ccharg\ is the correct expression since the K\"ahler cone 
of $X$ is generated by the noncompact divisors $D_1$ and $D_2$. 
\enddel
In terms of the pure brane basis $\tilde S$ the large volume central
charge is 
\eqn\lrlper{
Z^{\rm lv}(t_1,t_2;\tilde S)=
-\int_Xe^{-(t_1D_1+t_2D_2)}{\rm ch}(\tilde S)\sqrt{{\rm Td}(X)}=
\left[\matrix{
-{1\over 2}t_1^2+{3\over 2}t_1-{5\over 4} \cr
-{1\over 2}(3t_2^2+2t_1t_2)+t_1+{5\over 2}t_2-{7\over 6} \cr
-1 \cr
t_1-1 \cr
t_2-1 \cr
}\right].}

We will now discuss monodromy by assuming that the
assertions made in this section are correct.
The comparison with the mirror geometry is rather technical and can be
found in appendix A.
We want to find monodromy matrices acting on the charge vectors,
$n\to n\cdot M$, such that 
$n \cdot Z(\tilde S) = n\cdot M\cdot Z_{\rm mt}(\tilde S)$, where 
$Z_{\rm mt}(\tilde S)=M^{-1}Z(\tilde S)$ is the monodromy transformed
version of $Z$.
The monodromy around the orbifold locus cyclically permutes the charge
vectors \nch . Therefore, we obtain: 
\eqn\orbmonodromy{
M_{\rm orb}=\left[\matrix{
-2& 0 & 1 & 1 & 0 \cr
-2& 0 & 0 & 1 & 1 \cr
 0& 0 & 1 & 0 & 0 \cr
-2& 1 & 1 & 0 & 0 \cr
-1& -2& 0 & 2 & 1 
}\right]
}
Also, we can easily compute the large radius limit monodromies,
$M_{t_1}$ and $M_{t_2}$. On the 
sheaves defined on the exceptional divisors the actions of the
monodromies come from tensoring with the restrictions of $\co_X(D_1)$
and $\co_X(D_2)$. Therefore, the large radius limit monodromy
$M_{t_1}$ 
acts as following: the exceptional collection $\{\co_{\IP^2}(-n), {\cal V},
\co_{\IP^2}(-n+1)\}$  on $\IP^2$ is mutated to another exceptional 
collection, 
$\{\co_{\IP^2}(-n+1), \widetilde{\cal V}, \co_{\IP^2}(-n+2)\}$, while
on $\IF_3$ is given by the tensoring with $\co_{\IF_3}(f)$, therefore 
taking regular exceptional pairs into regular exceptional pairs.  
\par The action of the monodromy $M_{t_2}$ is represented by tensoring
any sheaf supported on $\IF_3$ with $\co_{\IF_3}(h+3f)$, hence again
transforming regular pairs into regular pairs, while leaving
any sheaf supported on $\IP^2$ invariant. 
\par Using \foumuknc\ it is possible to compute the action of the
monodromy around the principal component of the discriminant on the
generators of $K(X)$: $R_i^-$ with $i=1,\ldots ,4$ are invariant under
this transformation, but $R_0^-\mapsto \co_X(-2D_1-D_2)$. With the help
of \rstilde , we readily obtain the monodromy around the conifold locus:
\eqn\conmonodromy{
M_{\rm con}=\left[\matrix{
2 & 1 & 0 &-1 & 0 \cr
1 & 2 & 0 &-1 & 0 \cr
0 & 0 & 1 & 0 & 0 \cr
2 & 2 & 0 &-1 & 0 \cr
1 & 1 & 0 &-1 & 1
}\right]
}
The conifold monodromy, although preserving
exceptional collections, acts in a very different way on
$K^c(X)$. For example, we have $-[\co_{{\IF}_3}(-h)]\mapsto
[\co_{{\IP}^2}]$ and $[\co_{{\IP}^2}(-1)]\mapsto -[\co_{{\IF}_3}]$, that is
the D4-branes can 'jump' from one exceptional divisor to
another. However, as remarked in \aspmon , this is not very surprising
since autoequivalences of $D^{b}(X)$ need not preserve the D-branes.

\newsec{Beyond $\IC^d/\IZ_n$}

Up to now we have only considered cases of the type $\IC^d/\IZ_n$ with
a single triangulation. 
We now want to examine the range of validity of our statements
regarding the $S_i^-$ and monodromy.
We first present another example, the resolution of 
$\IC^3/(\IZ_2\times\IZ_2)$, which is still an orbifold but has several 
interesting features: 
It is not of the simple $\IZ_n$ type, it allows for more than one
triangulation, its resolution
involves three new non-compact toric divisors but no compact toric
divisor, and finally it is a three parameter model whose GKZ system
can be solved explicitly.
We will be able to show explicitly that the $S_i^-$ vanish at
(branches of) the principal component of the discriminant locus and
nowhere else. 
Aspects of D-brane states on this model have been studied previously 
in e.g. \BRG\AP .
Finally we examine the possibility of extending our results to cases
not of the McKay type.
We find that they still hold in many examples but not in general.

\noindent {\bf Example 4:}\hbr
{\it A toric resolution of} $\IC^3/(\IZ_2\times \IZ_2)$: 
A singular space of the type $\IC^3/(\IZ_2\times \IZ_2)$ where every
non-trivial element of $\IZ_2\times \IZ_2$ acts by flipping the sign
of two of the three coordinates of $\IC^3$ can be resolved by
introducing three additional non-compact divisors and three compact
curves.
There are several distinct possibilities for choosing the curves.
\ifig\fancthztwztw{$G$-Hilb resolution of $\IC^3/\IZ_2\times \IZ_2$.}
{\epsfxsize2in\epsfbox{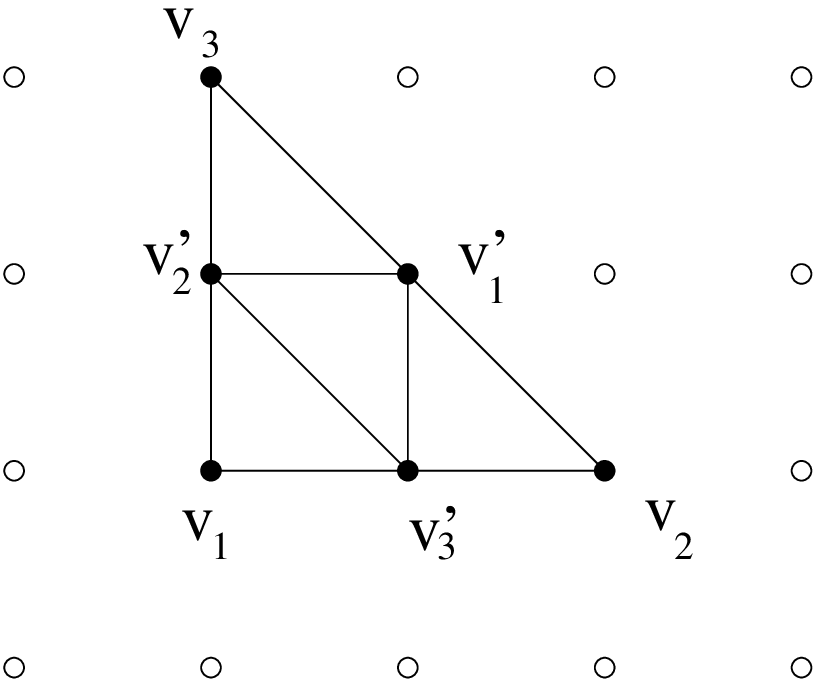}}
We use the $G$-Hilb resolution depicted in \fancthztwztw.
The Mori cone is generated by the following vectors:
\eqn\ckyzns{\eqalign{
l^{(1)}&=(1,~~0,~~0,~~1,~-1,~-1),\cr
l^{(2)}&=(0,~~1,~~0,~-1,~~1,~-1),\cr
l^{(3)}&=(0,~~0,~~1,~-1,~-1,~~1).}}
The generators of the K\"ahler cone are the divisors $D_1$, $D_2$ and
$D_3$ corresponding to the vanishing of the coordinates of $\IC^3$.
%The tautological line bundles forming the basis of $K(X)$ are
%given by
%\eqn\thersgenns{R_0=\co_X,~~R_i=\co_X(-D_i),~i=1,\ldots ,3}
The mirror geometry is determined by 
\eqn\cthztwpol{
a_1x_1^2+a_2x_2^2+a_3x_3^2+a_1'x_2x_3+a_2'x_1x_3+a_3'x_1x_2=0.}
The large complex structure coordinates $z_i$ are 
\eqn\cthztwlcsc{
z_1={a_1a_1'\over a_2'a_3'},~~~z_2={a_2a_2'\over a_1'a_3'},~~~
z_3={a_3a_3'\over a_1'a_2'}}
and the orbifold coordinates are 
\eqn\cthztworbc{
\phi_1={a_1'\over\sqrt{a_2a_3}}={1\over\sqrt{z_2z_3}},~~~~
\phi_2={a_2'\over\sqrt{a_1a_3}}={1\over\sqrt{z_1z_3}},~~~~
\phi_3={a_3'\over\sqrt{a_1a_2}}={1\over\sqrt{z_1z_2}}.}
The principal component of the discriminant locus is determined by
\eqn\cthztwdisc{
4 z_1 z_2 z_3 -z_1-z_2-z_3+1=0.}
The simplest formulation of the GKZ system can be obtained by mixing
large complex structure and orbifold coordinates.
We find the operator
\eqn\cthztwgkz{
{\cal D}_1=(\Theta_{\phi_2}+\Theta_{\phi_3})\Theta_{\phi_1}
-2z_1\Theta_{\phi_2}\Theta_{\phi_3}}
from the first Mori cone vector and the same operator with cyclically
permuted indices for the other two Mori cone vectors.
This simply implies
\eqn\mixedvan{\Theta_{\phi_2}\Theta_{\phi_3}\Pi=
\Theta_{\phi_1}\Theta_{\phi_3}\Pi=
\Theta_{\phi_1}\Theta_{\phi_2}\Pi=0} 
for any solution $\Pi$, i.e. there must be a basis of solutions
depending only on at most one of the $\phi_i$.
The sums of two Mori cone vectors lead to operators of the type
\eqn\cthztwgkzs{
{\cal D}_1'=
(\Theta_{\phi_1}+\Theta_{\phi_3})(\Theta_{\phi_1}+\Theta_{\phi_2})
-4z_1z_2\Theta_{\phi_1}(\Theta_{\phi_1}-1),}
which upon using \cthztworbc\ and \mixedvan\ implies
\eqn\cthztwgkzt{
\left((1-4\phi_1^{-2})\Theta_{\phi_1}+4\phi_1^{-2}\right)
\Theta_{\phi_1}\Pi=0.}
The whole GKZ system has three solutions of the type
$\ln((\phi_i+\sqrt{\phi_i^2-4})/2)$ and, as always, a constant solution.
Upon returning to large complex structure variables, we obtain 
\eqn\cthztwsol{
\Pi_0=1,~~~\Pi_1=\ln\left({1+\sqrt{1-4z_2z_3}\over 2}\right)-
{1\over 2}(\ln z_2+\ln z_3)}
and the corresponding index-permuted expressions for $\Pi_2$ and $\Pi_3$.

The divisors $F_i$ determining the line bundles $R_i$ are just $D_1$,
$D_2$ and $D_3$ and we find 
\eqn\cthztwS{
{\rm ch}^c(S^-_0)=p+C_1+C_2+C_3,~~~{\rm ch}^c(S^-_1)=-C_1,~~~
{\rm ch}^c(S^-_2)=-C_2,~~~{\rm ch}^c(S^-_3)=-C_3}
where $C_1$ is the compact curve at the intersection of $D_2'$ and
$D_3'$ etc.
In terms of structure sheaves we can represent $S^-_0$ as
$\tilde S_{C_1}+\tilde S_{C_2}+\tilde S_{C_3}-2\tilde S_p$.
Noticing that all three curves intersect in the same point, we find
that we can again view $S^-_0$ as the object whose restriction to any
compact toric cycle is the structure sheaf of that cycle.

The central charges are determined by $Z^{\rm lv}(t;S^-_0)=-1+t_1+t_2+t_3$
and $Z^{\rm lv}(t;S^-_i)=-t_i$ for $i\in\{1,2,3\}$, leading to
\eqn\cthztwcc{
Z(S^-_0)= 2-{\Pi_1+\Pi_2+\Pi_3\over 2\pi i},~~~ 
Z(S^-_1)=-1+{-\Pi_1+\Pi_2+\Pi_3\over 2\pi i},~~~\hbox{etc.}}

At the orbifold point $\phi_1=\phi_2=\phi_3=0$ we have the following
situation: 
The moduli space develops a $\IZ_2\times\IZ_2$ singularity.
Provided we make the right choice of sheets for the square roots and
logarithms, we find  $\Pi_1=\Pi_2=\Pi_3=3\pi i/2$ and thus
$Z(S^-_i)=-1/4$ for any $i$.
The `orbifold monodromy' $\phi_2\to -\phi_2$, $\phi_3\to -\phi_3$ acts as 
\eqn\cthztwom{
\Pi_2\leftrightarrow 3\pi i-\Pi_2,~~\Pi_3\leftrightarrow 3\pi i-\Pi_3,~~~~
S^-_0\leftrightarrow S^-_1,~~S^-_2\leftrightarrow S^-_3,}
and the other elements of the orbifold monodromy act in similar ways.

$S^-_0$ can become massless only if 
\eqn\massless{
(1+\sqrt{1-4z_2z_3})(1+\sqrt{1-4z_1z_3})(1+\sqrt{1-4z_1z_2})=8z_1z_2z_3.}
We can rewrite this in the form 
$\sqrt{1-4z_2z_3}~({\rm E}_1 )=({\rm E}_2 )$
such that E$_1$ and E$_2$ are expressions that do not contain
$\sqrt{1-4z_2z_3}$.
Then a necessary condition for \massless\ to hold is 
$(1-4z_2z_3)~({\rm E}_1 )^2=({\rm E}_2 )^2$ and we can proceed to
eliminate the other square roots in the same way.
The result is an equation proportional to the square of the expression
determining the principal component of the discriminant locus
\cthztwdisc.
Conversely, if we solve \cthztwdisc, e.g. by setting 
$z_1=(1-z_2-z_3)/(1-4z_2z_3)$, plug this into \cthztwsol\ and choose
the right sheets, we find that
$Z(S^-_0)$ indeed vanishes.
The same type of analysis works for the other $S^-_i$.

\bigskip
At this point it is natural to ask whether the analog of $S_0^-$,
i.e. the sheaf that is equal to the structure sheaf upon restriction
to any compact toric cycle but of rank zero away from these cycles,
might lead to monodromies in cases that are not related to McKay
correspondence.
It turns out that this is very often the case (at least for
sufficiently simple examples), but not true in general.

\ifig\nonmckay{Examples not of the McKay type.}
{\epsfxsize3in\epsfbox{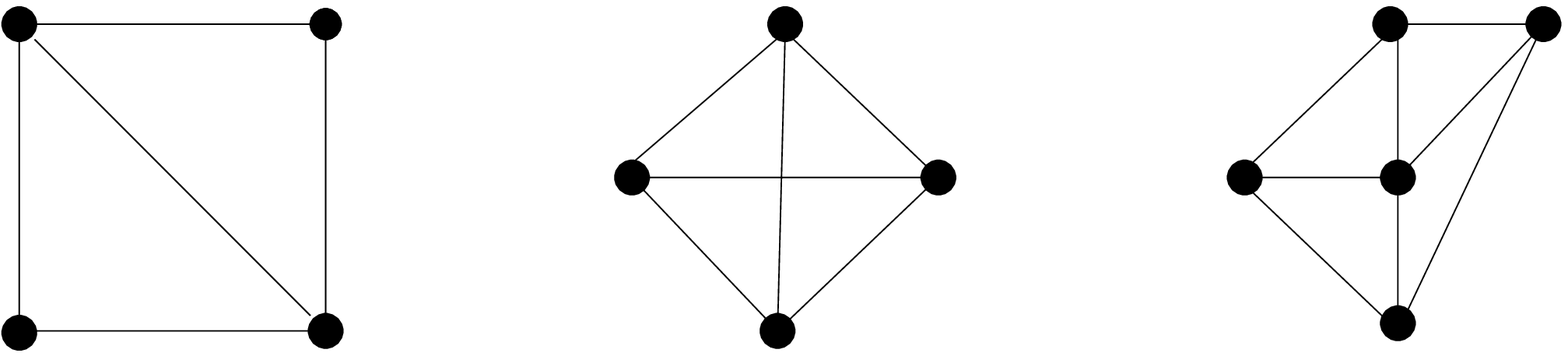}} 
Three examples where this works are shown in \nonmckay.
The first of these is the resolution of a conifold singularity and
exactly solvable. 
The other two (anticanonical line bundles over $\IF_0\simeq
\IP_1\times \IP_1$ and $\IF_1$, respectively) are two parameter models
that we treated with analyses similar to the ones used for example 3 
($\IC^3/\IZ_5$). 

\ifig\halfhexagon{Symmetric triangulation of the half-hexagon.}
{\epsfxsize2in\epsfbox{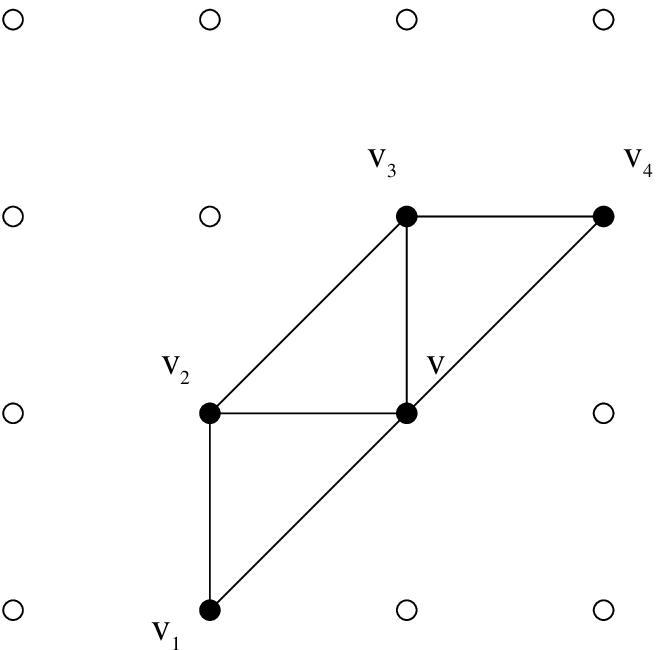}} 

As a counterexample, consider the Calabi-Yau manifold depicted in
\halfhexagon , whose GKZ system is again solvable. 
Here we find the following: If we choose as the line bundles $R_i$ the
ones determined by the generators of the K\"ahler cone, we still find
that the corresponding $S_0^-$ has the same restriction to compact
toric cycles as $\co_X$.
However, only two of the three generators of $K^c(X)$ become massless
at the conifold locus (these statements are true for any triangulation). 
In particular, for the symmetric triangulation the
vanishing locus of the central charge of $S_0^-$ does not coincide with
the conifold locus.

\bigskip
\noindent{\it Acknowledgements}

We would like to thank Philip Candelas and Duiliu Diaconescu for very
useful conversations.

\appendix{A}{Comparison of GKZ solutions and K-theory results for 
$\IC^3/\IZ_5$}
\noindent
The GKZ operators\foot{This GKZ system has also been studied in \mukray
.} corresponding to the Mori cone generators \chvect\
are given by 
\eqn\gkza{
{\cal D}^{(1)}=\partial_{a_1}\partial_{a_3}\partial_{a_5}
               -\partial_{a_4}^3,~~~
{\cal D}^{(2)}=\partial_{a_2}\partial_{a_4}-\partial_{a_5}^2.}
This can be turned into a system involving $z_1$, $z_2$ by standard
manipulations described above.
In this way we arrive at the following expressions in terms of
$\Theta_{z_i}:=z_i{\partial\over\partial z_i}$: 
\eqn\gkz{\eqalign{
&{\cal D}_1=\Theta_{z_1}^2(\Theta_{z_1}-2\Theta_{z_2})
-(\Theta_{z_2}-3\Theta_{z_1}+1)(\Theta_{z_2}-3\Theta_{z_1}+2)
(\Theta_{z_2}-3\Theta_{z_1}+3)z_1 \cr
&{\cal D}_2=\Theta_{z_2}(\Theta_{z_2}-3\Theta_{z_1})-
(\Theta_{z_1}-2\Theta_{z_2}+1)(\Theta_{z_1}-2\Theta_{z_2}+2)z_2, }}
Solutions to this system can be obtained by considering
\eqn\fundper{\eqalign{
\Pi(z_1,z_2;\rho_1,\rho_2):=
\sum_{n_1,n_2=0}^{\infty}
&z_1^{n_1+\rho_1}z_2^{n_2+\rho_2}
{\Gamma(1+\rho_1)^2\over\Gamma(1+n_1+\rho_1)^2}
{\Gamma(1+\rho_2)\over\Gamma(1+n_2+\rho_2)}
\cr &
\times{\Gamma(1+\rho_1-2\rho_2)\over\Gamma(1+n_1-2n_2+\rho_1-2\rho_2)}
{\Gamma(1+\rho_2-3\rho_1)\over\Gamma(1+n_2-3n_1+\rho_2-3\rho_1)}
\cr }}
(the coefficients of $n_i$, $\rho_i$ in the $\Gamma$--functions are
the entries of the Mori cone vectors \chvect) and its partial
derivatives w.r.t the $\rho_i$ at $\rho_1=\rho_2=0$. 
We use:
\eqn\Gam{
{\Gamma(1+\rho)\over\Gamma(1+n+\rho)}=
\matrix{1 & n=0\cr
        [(1+\rho)(2+\rho)\cdots (n+\rho)]^{-1} & n>0\cr  
        \rho (\rho-1)\cdots (\rho+n+1)  & n<0\cr}}
\eqn\Gamrho{
{\partial\over\partial\rho}
\left({\Gamma(1+\rho)\over\Gamma(1+n+\rho)}\right)_{\rho=0}=
\matrix{0 & n=0\cr
        -S_n/n! & n>0\cr  
        (-1)^{-n-1}(-n-1)! & n<0\cr}}
\eqn\Gamrhorho{
{\partial^2\over\partial\rho^2}
\left({\Gamma(1+\rho)\over\Gamma(1+n+\rho)}\right)_{\rho=0}=
\matrix{0 & n=0\cr
        {\rm finite} & n>0\cr  
        2(-1)^{-n}S_{-n-1}(-n-1)!& n<0\cr}}
where $S_n=1+1/2+\ldots +1/n$.
This yields the constant solution $\Pi(z_1,z_2;0,0)=1$ and, with 
$\Pi_{i_1\cdots i_k}$ for $(\partial^k \Pi / 
\partial \rho_{i_1}\ldots \partial \rho_{i_k})|_{\rho=0}$,\hfill\smallskip\noindent
%\eqn\logone{
$\Pi_1=\ln z_1+\sum z_1^{n_1}z_2^{n_2}A_{n_1n_2}-
      3\sum z_1^{n_1}z_2^{n_2}B_{n_1n_2},$\hfill\smallskip\noindent
%}
%\eqn\logtwo{
$\Pi_2=\ln z_2-2\sum z_1^{n_1}z_2^{n_2}A_{n_1n_2}+
       \sum z_1^{n_1}z_2^{n_2}B_{n_1n_2},$\hfill\smallskip\noindent
%}
%\eqn\logoneone{\eqalign{&
$\Pi_{11}=
(\ln z_1)^2+2\ln z_1(\sum z_1^{n_1}z_2^{n_2}A_{n_1n_2}-
3\sum z_1^{n_1}z_2^{n_2}B_{n_1n_2})
-6\sum z_1^{n_1}z_2^{n_2}C_{n_1n_2}$\hbr
${}~~~~~~~~+\sum z_1^{n_1}z_2^{n_2}A_{n_1n_2}
(-2S_{2n_2-n_1-1}+6S_{n_2-3n_1}-4S_{n_1})$\hbr
${}~~~~~~~~+\sum z_1^{n_1}z_2^{n_2}B_{n_1n_2}
(-18S_{3n_1-n_2-1}+6S_{n_1-2n_2}+12S_{n_1})$,\hfill\smallskip\noindent
%\cr
%}}
%\eqn\logonetwo{\eqalign{&
$\Pi_{12}=\ln z_1\ln z_2+\ln z_1(-2\sum z_1^{n_1}z_2^{n_2}A_{n_1n_2}+
\sum z_1^{n_1}z_2^{n_2}B_{n_1n_2})$\hbr
${}~~~~~~~~+\ln z_2(\sum z_1^{n_1}z_2^{n_2}A_{n_1n_2}
-3\sum z_1^{n_1}z_2^{n_2}B_{n_1n_2})
+7\sum z_1^{n_1}z_2^{n_2}C_{n_1n_2}$\hbr
${}~~~~~~~~+\sum z_1^{n_1}z_2^{n_2}A_{n_1n_2}
(4S_{2n_2-n_1-1}-7S_{n_2-3n_1}+4S_{n_1}-S_{n_2})$\hbr
${}~~~~~~~~+\sum z_1^{n_1}z_2^{n_2}B_{n_1n_2}
(6S_{3n_1-n_2}-7S_{n_1-2n_2}-2S_{n_1}+3S_{n_2}),$\hfill\smallskip\noindent
%}}
%\eqn\logtwotwo{\eqalign{&
$\Pi_{22}=(\ln z_2)^2+2\ln z_2(-2\sum z_1^{n_1}z_2^{n_2}A_{n_1n_2}+
\sum z_1^{n_1}z_2^{n_2}B_{n_1n_2})
-4\sum z_1^{n_1}z_2^{n_2}C_{n_1n_2}$\hbr
${}~~~~~~~~+\sum z_1^{n_1}z_2^{n_2}A_{n_1n_2}
(-8S_{2n_2-n_1-1}+4S_{n_2-3n_1}+4S_{n_2})$\hbr
${}~~~~~~~~+\sum z_1^{n_1}z_2^{n_2}B_{n_1n_2}
(-2S_{3n_1-n_2-1}+4S_{n_1-2n_2}-2S_{n_2}),$\hfill\smallskip\noindent
%}}
where\hfill\smallskip\noindent
%\eqn\ABC{\eqalign{&
$A_{n_1n_2}={(2n_2-n_1-1)!\over (n_2-3n_1)!(n_1!)^2n_2!}(-1)^{2n_2-n_1-1},$
\hfill\smallskip\noindent
$B_{n_1n_2}={(3n_1-n_2-1)!\over (n_1-2n_2)!(n_1!)^2n_2!}
(-1)^{3n_1-n_2-1},$ 
\hfill\smallskip\noindent
$C_{n_1n_2}={(2n_2-n_1-1)!(3n_1-n_2-1)!\over 
(n_1!)^2n_2!}(-1)^{2n_2-n_1-1+3n_1-n_2-1}$\hfill\smallskip\noindent
and the summations are taken over those values of $n_1$, $n_2$ where
the arguments of all factorials are non-negative.
Of the three expressions obtained by taking second derivatives only
the first one and the linear combination $3\Pi_{22}+2\Pi_{12}$ of the
other two actually solve the GKZ system \gkz .
We note that there is also a linear combination of third derivatives
(involving third powers of logarithms) that is annihilated by both
operators occurring in \gkz .
The reason is that this system is not yet complete as we have written
it:
In principle we should write down a GKZ operator for every curve in
the Mori cone.
Taking as an additional charge vector the sum $l^{(1)}+l^{(2)}$ of our
Mori cone generators, we see that the triple-log solution is excluded.
\ifig\realmodsp{The real part of the discriminant locus.}
{\epsfxsize2.5in\epsfbox{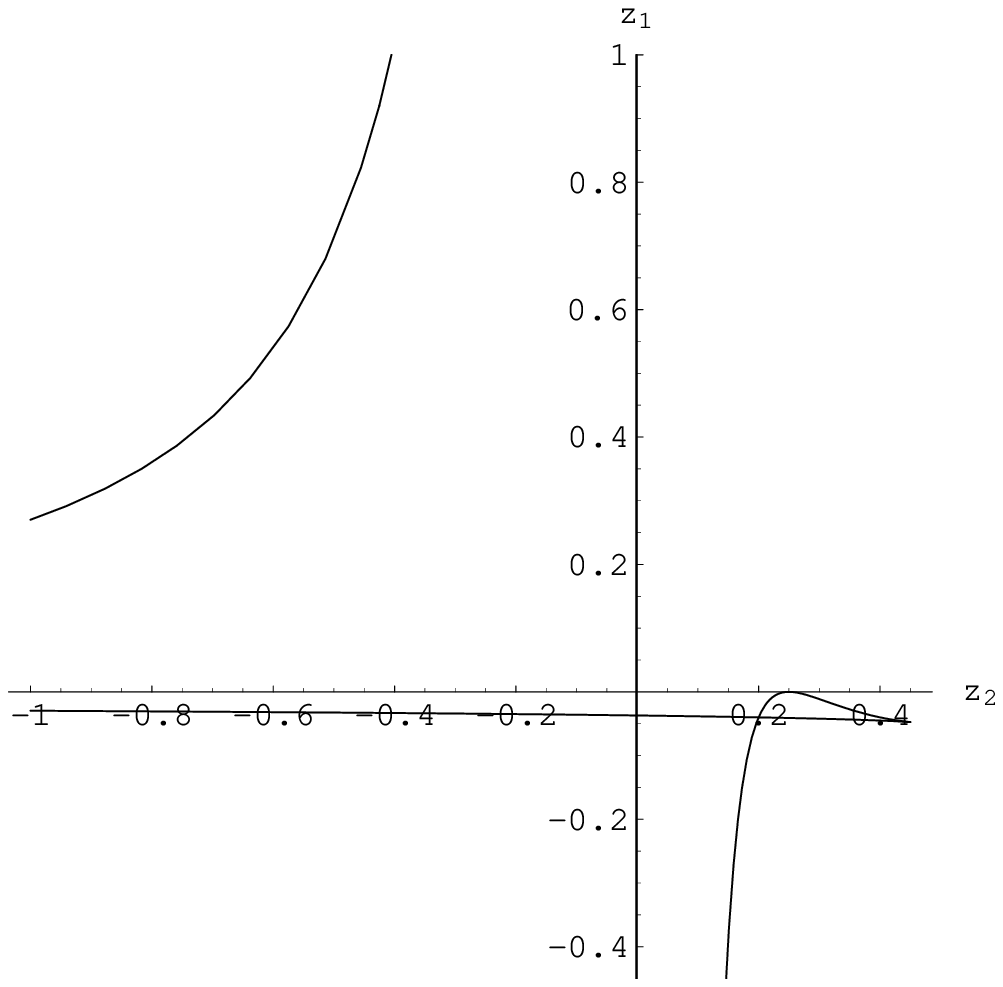}}

We now want to study monodromies of the GKZ solutions around the loci
where $\cal R$ degenerates.
This is an easy exercise for the divisors $z_1=0$, $z_2=0$ where the
monodromy is determined by $\ln z_i\to\ln z_i +2\pi i$.
We find that the following set of solutions is well behaved (i.e.,
transforms by an $SL(5,\IZ)$ matrix) under the monodromies 
$z_i\to e^{2\pi i} z_i$:
\eqn\lcsper{
\left[\matrix{
1 \cr
{1\over 2\pi i}\Pi_1 +{\rm const.}\cr
{1\over 2\pi i}\Pi_2 +{\rm const.}\cr
{1\over 2(2\pi i)^2}\Pi_{11}-{1\over 2(2\pi i)}\Pi_1+{\rm const.}\cr
{1\over 2(2\pi i)^2}(3\Pi_{22}+2\Pi_{12})-{1\over 2(2\pi i)}\Pi_2
+{\rm const.}\cr
\del
{1\over 2}t_1^2-{1\over 2}t_1+{1\over 4} + O(z_1,z_2)\cr
{1\over 2}(3t_2^2+2t_1t_2)-{1\over 2}t_2+{1\over 6} + O(z_1,z_2)\cr
\enddel
}\right].}
In the large complex structure coordinate patch the discriminant locus
consists of several different branches.
The slice through real $z_1$, $z_2$ is shown in \realmodsp.
There are two branches with $z_1\le 0$. 
The one with $z_2>0$ is tangent to the axis $z_1=0$ in $z_2=1/4$. 
Parts of this branch are at the boundary of the domain of
convergence of the $\Pi$'s in such a way that there is
convergence like $1/n^2$.
Through a numerical analysis we found that at this locus 
\eqn\abranch{
-{1\over 2}(2\Pi_{12}+3\Pi_{22})+{1\over 2}\Pi_2-{1\over 6}=0,}
which corresponds to the vanishing of $Z(S_4^-)$ if we make the
identifications $t_1\sim 1+\ln z_1$, $t_2\sim \ln z_2$.
This is equivalent to the vanishing of a $z_2$-monodromy transformed
version of $S_4^-$ with \ident.
Similarly we find at the other branch with $z_1<0$ which
intersects $z_2=0$ at $z_1=-1/27$ that 
\eqn\bbranch{
-{1\over 2}\Pi_{11}+{1\over 2}\Pi_1-{1\over 4}=0.}
This corresponds to a $z_1$-monodromy transformed version of $S_2^-$
vanishing.
The third branch, with $z_1>0$ and $z_2<0$ is beyond the region of
convergence.
We have preliminary evidence that at this branch a $z_2$-monodromy
transformed version of $S_0^-$ becomes massless.

\listrefs
\end